\documentclass[11pt]{article}
\usepackage{graphicx}
\usepackage{longtable}
\usepackage[utf8]{inputenc}
\usepackage{geometry}
\usepackage{tabularx}
\usepackage{helvet}         
\usepackage{graphicx}
\usepackage{amsmath, amssymb}
\usepackage{hyperref}
\usepackage{url}
\usepackage{titlesec}
\usepackage{fancyhdr}
\usepackage{newunicodechar}
\newunicodechar{₂}{$_2$}
\usepackage{cite}
\usepackage{enumitem}
\usepackage{titling}
\usepackage{booktabs}
\usepackage{afterpage}
\usepackage{acronym} 
\usepackage{xcolor}
\usepackage{setspace}
\usepackage{tikz}
\usetikzlibrary{shapes.geometric, arrows.meta, positioning, fit}

\usepackage{acronym}

\definecolor{fortissblue}{RGB}{0,70,150}
\titleformat{\section}{\color{fortissblue}\normalfont\Large\bfseries}{\thesection}{1em}{}
\titleformat{\subsection}{\color{fortissblue}\normalfont\large\bfseries}{\thesubsection}{1em}{}
\titleformat{\subsubsection}{\color{fortissblue}\normalfont\normalsize\bfseries}{\thesubsubsection}{1em}{}
\begin{document}

\begin{titlepage}
    \vspace*{3cm}
    {\color{fortissblue}\Huge\bfseries Towards Human-Centric and Sustainable 6G Services \par}
    \vspace{2cm}
    {\color{fortissblue}\LARGE The fortiss Research Perspective\par}
    \vspace{1cm}
    {\color{fortissblue}White Paper v1.0\par}
    
    \vfill

    \vspace{0.5cm}
    {\color{fortissblue}\Large fortiss GmbH, Munich, Germany \par}
    \vspace{0.5cm}
    {\color{fortissblue}\large \today \par}
    \vfill
    \includegraphics[width=0.3\textwidth]{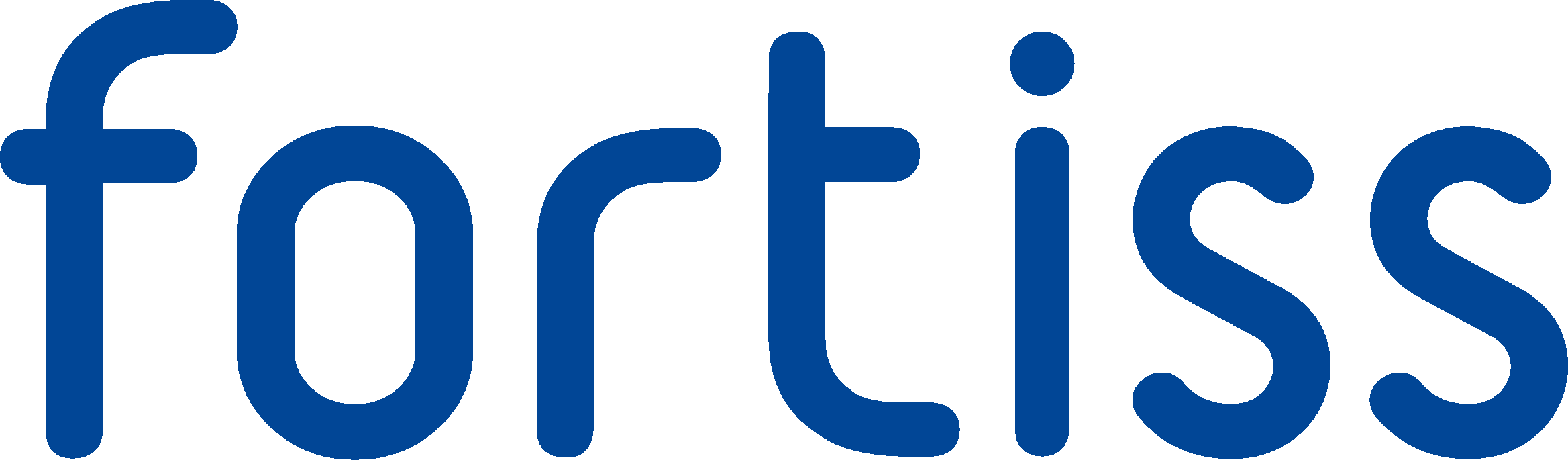} 
    \vspace*{2cm}
\end{titlepage}


\section*{Executive Summary}
As a leading research institute in software-intensive systems, \textbf{fortiss} is at the forefront of shaping the \ac{6G} vision. Our mission is to ensure that 6G technologies and architectures not only meet technical benchmarks but are also grounded in societal needs. fortiss actively contributes to national and international 6G initiatives, including standardization bodies and collaborative R\&D programs. We focus on advancing software-based, AI-enabled, and sustainable communication services that place human needs at the core of innovation.
The sixth generation of mobile communications (6G) is set to redefine digital connectivity by integrating cognitive intelligence, decentralized orchestration, and sustainability-driven architectures. As demand grows for ultra-reliable, low-latency communication (\ac{URLLC}) and user-driven digital experiences, 6G must surpass its predecessors.
It will do so by leveraging AI-native networking, Edge–Cloud resource orchestration, and energy-efficient data frameworks. These innovations must balance technical performance with societal and individual needs.

This white paper outlines fortiss’ vision for a human-centric, sustainable, and AI-integrated 6G network. It presents core research areas, including semantic communication, green orchestration, and distributed AI, and links them to societal and technological challenges. The vision considers a system architecture integrating Edge-Cloud-IoT intelligence with collaborative and ethical principles, supported by ongoing standardization and research efforts. The document concludes with strategic enablers and open research questions to foster community collaboration toward the 2030 6G vision.


The white paper is intended for a broad audience of stakeholders involved in the future of communications, e.g., including researchers, industry leaders, policymakers, and system developers. Its purpose is to outline the strategic vision and ongoing contributions of fortiss to the evolution of 6G technologies, emphasizing human-centric design, AI integration, and sustainable network architectures. By highlighting both challenges and research directions, the paper aims to foster collaboration and inspire innovation across disciplines.
\vspace{1em}
\\
\textbf{Invitation to Collaborate:}  
fortiss invites industry partners, academic institutions, and public sector organizations to join us in defining the future of 6G. Whether through joint research projects, testbed development, policy engagement, or standardization efforts, we welcome opportunities to co-create a 6G ecosystem that is not only technologically advanced but also ethically grounded and inclusive by design.

\vspace*{1cm}

\textbf{\color{fortissblue}Keywords:} 6G, AI, Edge-Cloud orchestration, human-centricity, software engineering.

\newpage
\tableofcontents
\newpage


\section{Introduction}
The vision for the \textit{sixth generation (6G)} of wireless and mobile communications is currently taking shape worldwide through a variety of initiatives. At its core, 6G represents the next phase in the evolution of cellular networks, a progression that began with 2G and 3G  primarily enabling voice and text communication between individuals. With 4G, the focus shifted significantly, driven by the widespread adoption of mobile technologies for high-volume voice and video consumption. The advent of \ac{5G} introduced support for the \textit{Internet of Things (IoT)} and enabled greater automation, leading to a surge in data usage and service consumption across the Internet~\cite{5g_evolution, industryautomation}. 
The widespread adoption of 5G is largely driven by its ability to support a broad range of services and applications centered around \textit{\ac{UE}}, i.e., personal devices such as smartphones, tablets, and other connected gadgets. A key shift from 4G to 5G lies in this enhanced support for user-centric services, enabling more personalized, responsive, and data-intensive experiences.
6G is expected to go beyond the incremental evolution of previous cellular generations, the so-called "Gs", by fundamentally redefining how networks interact with users, devices, and digital ecosystems~\cite{pennanen20246g}. This transformation is powered by advancements in software-defined networking, virtualization, edge–cloud computing, and \textit{\ac{AI}}. The convergence of these enablers is already paving the way for a new wave of human-centric applications and services, marking a profound transformation in network design and capability.
One of the significant changes in 6G relates with the massive deployment of IoT. According to Statista, global IoT connections are projected to rise from 18 billion in 2024 to over 32.1 billion by 2030\footnote{\url{ https://www.statista.com/statistics/1183457/iot-connected-devices-worldwide/}}. The consumer segment currently accounts for approximately 60\% of the overall market, a trend that is expected to continue into the next decade. This exponential growth reinforces the need for a robust, intelligent, and highly scalable 6G infrastructure.
However, realizing the 6G vision demands an interdisciplinary approach that extends beyond advancements in telecommunications alone. By 2030, 6G is expected to act as a catalyst for socially impactful use cases, enabling seamless data exchange and cognitive interaction between humans and cyber-physical systems. Future 6G applications are already being envisioned to go beyond traditional \textit{\ac{AR}/\ac{VR}}, incorporating multi-sensory experiences and fully immersive digital interactions~\cite{alsamhi2024multisensory}. Consequently, sensing, data communications, and AI will serve as foundational technological enablers of the 6G era\footnote{\url{https://digital-strategy.ec.europa.eu/en/policies/6g}}.
Achieving a holistic 6G vision requires cross-disciplinary brainstorming among communication engineers, computer scientists, social scientists, and industry stakeholders. It is essential to integrate existing infrastructure with forward-looking innovations while addressing emerging societal and ethical considerations. To this end, global research and innovation initiatives are actively working on 6G roadmaps to guide the first pilot deployments expected around 2030.
\textbf{As a leading research institute, fortiss is actively engaged in a broad range of research and development initiatives that contribute to shaping the next generation of wireless communication, i.e., 6G. }These efforts are particularly focused on enabling sustainable, secure, and human-centric services that are powered by software-defined and AI-driven network architectures. Recognizing that 6G represents more than just a technological leap, fortiss is approaching the challenge from an interdisciplinary perspective, integrating insights from computer science, engineering, human-computer interaction, and socio-technical systems.
The white paper aims to shed light in the multifaceted 6G research activities currently underway at fortiss. It highlights the institute’s unique contributions to the development of software-based enablers for 6G, the exploration of future use cases, and the creation of frameworks that prioritize human needs and values. Furthermore, the paper outlines fortiss’s strategic positioning within both the national and international 6G research landscape, demonstrating its commitment to collaboration, innovation, and impact across industry, academia, and policy.

As a hub for applied research and development in Bayern, Germany, fortiss contributes to shaping the 6G ecosystem by developing research across the following 6G areas of action:
\begin{itemize}
    \item \textbf{Research and Innovation in 6G Technologies}
    \subitem\textbf{\textit{Edge-Cloud continuum}}.  fortiss explores cognitive, decentralized edge-cloud orchestration techniques that optimize the management of 6G infrastructures. These advancements focus on user-centric applications, ensuring efficient data processing, computation, and networking.
    \subitem  {\textbf{\textit{ AI integration and development}}. fortiss pioneers AI-driven methodologies for data validation, network optimization, human-AI collaboration, context-aware AI, human-in-the-loop, and learning-based perception and control within 6G-connected environments.
    }
    \subitem  \textbf{\textit{Human-centric design}}.  We engineer data communication technologies that meet human needs, advancing universal accessibility, proactive inclusivity, digital wellbeing, and context-aware enhancement of \textit{\ac{QoE}}. 
    \subitem  \textbf{\textit{Software engineering}}.  We engineer data communication technologies that meet human needs, advancing universal accessibility, proactive inclusivity, digital wellbeing, and context-aware enhancement of QoE. 
     \item \textbf{Standardization and Collaboration}
     \subitem \textbf{Research to Standardization transfer}: fortiss actively participates in pursuing the transfer of technological research to standardisation.
    \subitem \textbf{Cross-Sector Collaboration}: fortiss engages with industries, academia, and government to align 6G advancements with practical needs.
\end{itemize}

\textbf{This white paper targets researchers, policymakers, and industry leaders involved in shaping the future of communication networks. It outlines fortiss’s strategic contributions to 6G development, focusing on AI integration, Edge–Cloud orchestration, semantic communication, and sustainability.
}

The paper is organized as follows. Section \ref{6G vision} introduces the 6G paradigm and its current status.
Section \ref{6Gresearch} presents the fortiss ongoing research contributions and activities towards the realization of 6G. Section \ref{fortiss6G} presents the positioning of fortiss in different international 6G activities.
Section \ref{6Gfutureinfortiss} debates on the realization of a 6G vision, based on the fortiss interdisciplinary research.
Section \ref{Conclusions} summarizes the white paper, highlighting main takeaways.

\newpage

\section{6G Vision, Enablers, and Challenges}
\label{6G vision}
\subsection{Vision}
The sixth generation of cellular communications (6G) is expected to launch commercially around 2030. Unlike its predecessors, 6G aims to meet the rising demands of data-intensive applications, ubiquitous connectivity, and ultra-low latency services. According to the \ac{ITU-R} Global 6G Vision (\ac{IMT}-2030) \footnote{\url{https://www.itu.int/en/mediacentre/Pages/PR-2023-12-01-IMT-2030-for-6G-mobile-technologies.aspx}}, six key use-case areas have been identified: \textbf{immersive communication}, \textbf{AI-integrated communication}, \textbf{hyper-reliable low-latency communication}, \textbf{ubiquitous and massive connectivity}, and \textbf{integrated sensing}.

These are underpinned by four core principles: \textbf{sustainability}, \textbf{connecting the unconnected}, \textbf{ubiquitous intelligence}, and \textbf{security/resilience}. To meet these goals, 6G will adopt a simplified, energy-efficient architecture designed to reduce latency and support trillions of interconnected devices—paving the way for widespread IoT applications across sectors such as Health, Energy, Manufacturing. 

\begin{figure}[h!]
\centering
\includegraphics[width=0.8\textwidth]{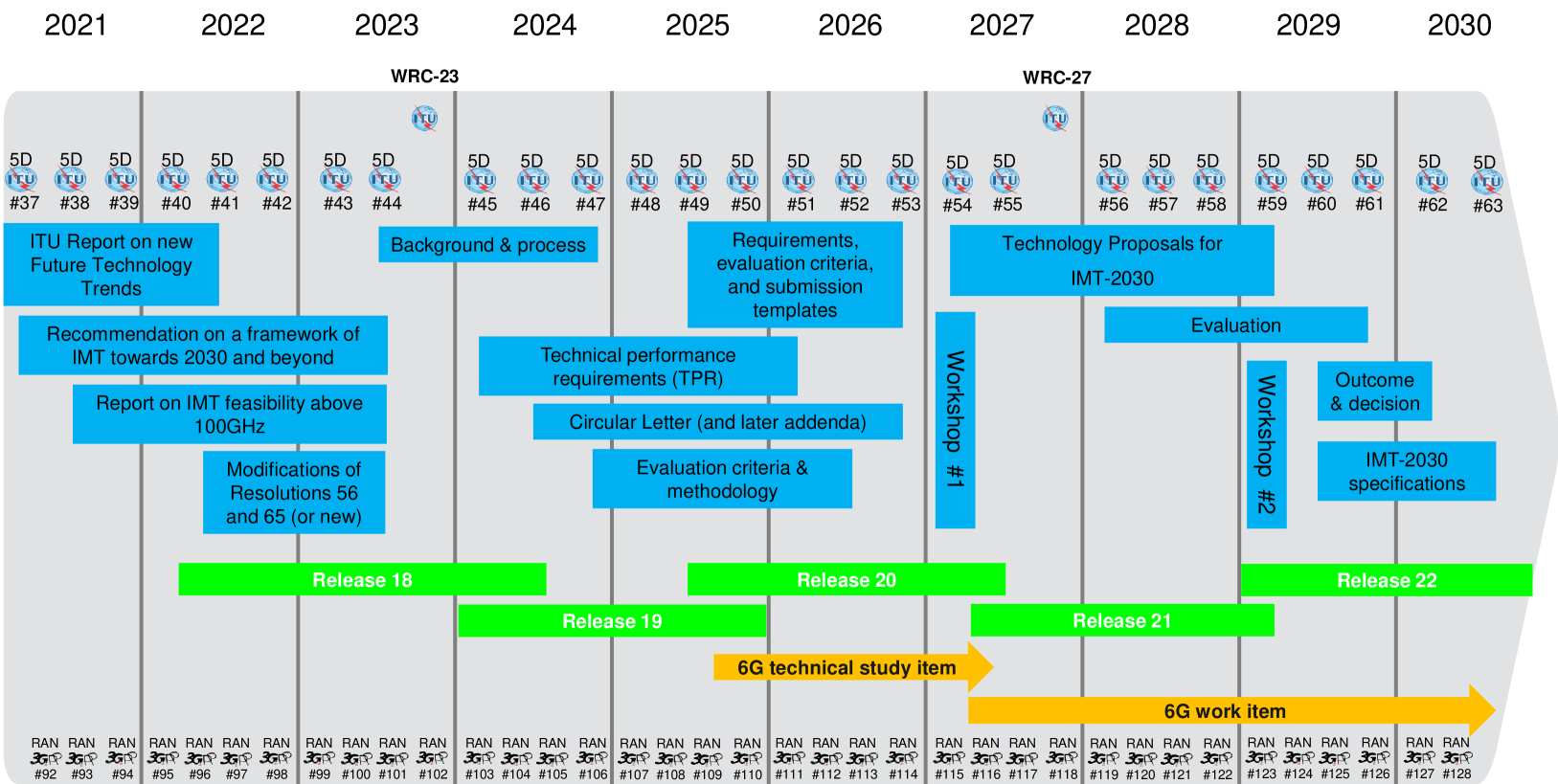}
\caption{\small \textit{Timeline for 6G studies. Blue boxes represent the ITU-R expected releases; green represent the proposed 3GPP releases; yellow represents the  3GPP phases}\cite{liu2025ituvisionframework6g}.}\label{fig:6Gtimeline}
\end{figure}

The timeline for 6G development, being steered by the \textit{3rd Generation Partnership Project }(3GPP)\footnote{\url{https://www.3gpp.org/}} and by the \textit{\ac{ITU}}\footnote{\url{https://www.itu.int/en/Pages/default.aspx}} and represented in Figure \ref{fig:6Gtimeline} is structured across multiple phases, with key milestones defining its evolution. 

The specification development and standardization process is expected to take place between 2025 and 2029, during which \textit{Standards Development Organizations (SDOs)} such as 3GPP, ITU, IEEE, will work towards defining 6G’s technical framework. Early-stage laboratory testing and pilot deployments are anticipated to begin around 2028, allowing for real-world validation of 6G technologies and their integration into existing communication infrastructures. A crucial milestone in this journey is the release of 3GPP Release 19, scheduled for completion in late 2025\footnote{\url{https://www.3gpp.org/specifications-technologies/3gpp-work-plan}}, which is laying the foundational groundwork for the development and deployment of 6G networks.
\subsection{Enablers, Perspective from the main 6G Ongoing Specifications}
The main 6G enablers are being debated and developed via multiple 3GPP releases, of which the main ones are release 19 (initiation of 6G Requirements studies); release 20 (Technical Studies on 6G Technologies); release 21 (First Set of 6G Technical Specifications). These are summarised in Table \ref{tab:3gpp_releases} and presented in the next sub-sections.

\begin{table}[h]
    \centering
    \caption{\textit{Main features in 3GPP Releases 19–21}.}
    \label{tab:3gpp_releases}
    \small
    \begin{tabular}{p{3.5cm} p{3.5cm} p{3.5cm} p{3.5cm}}
        \toprule
        \textbf{Feature} 
            & \textbf{19 5G-Advanced Start} 
            & \textbf{20 Finalizing 5G-Advanced} 
            & \textbf{21 1st 6G Release} \\
        \midrule
        \textbf{Spectrum} 
            & Sub-7 GHz \& mmWave 
            & New upper-mmWave bands 
            & Terahertz (THz) bands \\

        \addlinespace[0.5em]
        \textbf{AI/ML} 
            & Network optimization 
            & AI-driven automation 
            & AI-native 6G core \\

        \addlinespace[0.5em]
        \textbf{Extended Reality (XR)} 
            & Latency/scheduling improvements 
            & AI-based traffic steering 
            & Fully immersive, AI-enhanced \\

        \addlinespace[0.5em]
        \textbf{Non-Terrestrial Networks} 
            & IoT NTN, programmable payloads 
            & Voice/data over NTN, seamless handover 
            & Full 6G–NTN integration \\

        \addlinespace[0.5em]
        \textbf{Sustainability \& Energy} 
            & Power-saving modes, SSB optimizations 
            & Carbon footprint tracking, adaptive power 
            & Ultra-sustainable design \\

        \addlinespace[0.5em]
        \textbf{Security} 
            & 5G security updates 
            & AI-enhanced threat detection 
            & Quantum-resistant (\ac{PQC}) \\

        \addlinespace[0.5em]
        \textbf{6G Studies} 
            & Initial feasibility studies 
            & 6G requirements deep-dive 
            & First official 6G specifications \\
        \bottomrule
    \end{tabular}
\end{table}

\subsubsection{3GPP Release 19}
The 3GPP Release 19\footnote{\url{https://www.3gpp.org/specifications-technologies/releases/release-19}}, part of the ongoing 5G-Advanced evolution, is aimed at enhancing network performance, enabling new applications, and preparing the foundation for future 6G developments. In addition to its technical advancements, Release 19 includes initial studies on 6G requirements, helping to define the baseline for future specifications. The key enablers currently being addressed in this release include:
\begin{itemize}
    \item \textbf{Spectrum Management}. Key focus areas include spectrum diversity and mobility management. Release 19 enhances beamforming in the 6–7 GHz range with support for larger antenna arrays and introduces cost-efficient solutions for distributed \textit{Multiple-Input Multiple-Output (MIMO)}. It also improves mobility by accelerating beam management and extending Layer 1/2-triggered handovers between different 5G base stations e.g., \textit{next generation Node B, gNB}), supported by \ac{AI}/\ac{ML}-based predictions.
    \item \textbf{Support for advanced applications: Metaverse, \ac{XR}, Ambient power-enabled IoT~\cite{butt2024ambient}}. Release 19 enhances XR support through improved scheduling based on packet delay and reduced impact of measurement gaps. It also explores support for Ambient IoT devices powered by energy harvesting, targeting ultra-low power consumption and long lifecycles. Additionally, it introduces groundwork for localized mobile Metaverse services, with TR 22.856 defining related service requirements and frameworks\footnote{3GPP TR 22.856, \url{ https://portal.3gpp.org/desktopmodules/Specifications/SpecificationDetails.aspx?specificationId=4046}}, and TR 23.438 defining \textit{\ac{SEAL}} Digital assets facilitating application support for mobile metaverse services.
    
    \item \textbf{Network sustainability and automation}. The main focus is on energy efficient and AI/ML integration. Building upon previous enhancements, Release 19 introduces features such as on-demand \textit{\ac{SSB}} transmission in secondary cells and adaptive transmission periodicity to improve network energy efficiency. Furthermore, studies address energy efficiency and energy savings by design. In terms of AI/ML integration,  release 19 specifies a general framework for using AI/ML on the air interface, addressing use cases such as positioning and beam management, and continues exploring AI/ML applications in mobility and other areas.    
    \item \textbf{\ac{NTN}}. Release 19 improves uplink capacity for IoT, NTN, and NR-NTN, enabling more efficient device-to-satellite communication. It also introduces support for regenerative payloads, allowing satellites to handle on-board processing tasks. Downlink coverage is enhanced to ensure more reliable service, while energy efficiency is boosted through extended support for Reduced Capability (RedCap) devices in NTN, enabling low-power, cost-effective IoT applications~\cite{veedu2022toward}.
\end{itemize}
\subsubsection{3GPP Release 20}
Release 20\footnote{\url{https://www.3gpp.org/news-events/3gpp-news/sa-rel20}}, which is starting in 2025 and expected to be concluded in 2027 as the final part of the 5G-Advanced phase, focuses on technical studies for potential 6G technologies, e.g., advancements in radio interfaces and core network architectures to provide a smooth transition towards 6G technologies. Key focus is on a better integration and use of AI/ML, and on energy efficiency aspects, as well as on a better NTN integration. The main aspect being worked are as follows:
\begin{itemize}
\item \textbf{Network Energy Efficiency}. Release 20 emphasizes reducing the carbon footprint of communication services by introducing dynamic adjustments based on energy availability and enabling the reporting of carbon footprints for specific 6G services, e.g., fleets of mobile robots or vehicles, dense IoT.
\item \textbf{AI/ML integration}. Considers a more comprehensive integration of AI/ML for network management and other network operations, enhancing performance and user experience. This integration includes distributed learning and embedding AI within both network infrastructure and user devices.
\item \textbf{Support for advanced applications: \textit{XR/\ac{AR}}, Metaverse, A-IoT}. Release 20 focuses on \ac{QoS} handling and monitoring specifically addressing the needs of the advanced applications studied in release 19. Furthermore, for A-IoT release 20 expands the scope of \textit{\ac{A-IoT}} to include topologies better suited for cellular networks and potentially more capable devices, moving beyond limited \ac{RFID}-type functionalities.
\item \textbf{NTN integration}. Release 20 advances NTN capabilities by supporting low data rate voice communication over IoT-NTN connections, providing alternatives for satellite-based emergency calls, and enhancing mission-critical services through satellite access. 
\end{itemize}

\subsubsection{3GPP Release 21}
Release 21\footnote{\url{https://www.3gpp.org/news-events/3gpp-news/sa-rel20}} , expected to be released by the end of 2028, aims to provide the initial set of 6G technical specifications, serving as the basis for IMT-2030 submissions and paving the way for commercial 6G deployments anticipated around 2030. It will be the first official 6G standardization release and sets the overall 6G architecture, spectrum, and use-cases. Key enablers can be summarized as follows:
\begin{itemize}
\item \textbf{Advanced spectrum utilization}.  Release 21 is expected to support operations across a broader spectrum, including frequencies beyond those utilized in 5G, to accommodate higher data rates and new services.
\item \textbf{Enhanced Network Architecture}. The key focus is on  convergence of communication and sensing functionalities, enabling networks to not only transmit data but also perceive and interpret environmental contexts.
\item \textbf{AI/ML-driven Network Management}. Functionality to assist in a resilient integration of AI/ML in the context of network management and adaptive optimization, leading to more efficient and resilient networks.
\item \textbf{Sustainability and Energy-efficiency}. Emphasis is being placed in defining "Green networks", i.e., supporting and devising energy-efficient protocols and architectures to reduce the carbon footprint of mobile networks, aligning with global sustainability goals.
\item \textbf{Security and Trustworthiness}. The current focus is on the development of robust security mechanisms to protect against emerging threats, ensuring data integrity and user privacy in increasingly complex network environments. 
\end{itemize}

\subsection{Challenges}

\begin{figure}[h!]
\centering
\includegraphics[width=0.6\textwidth]{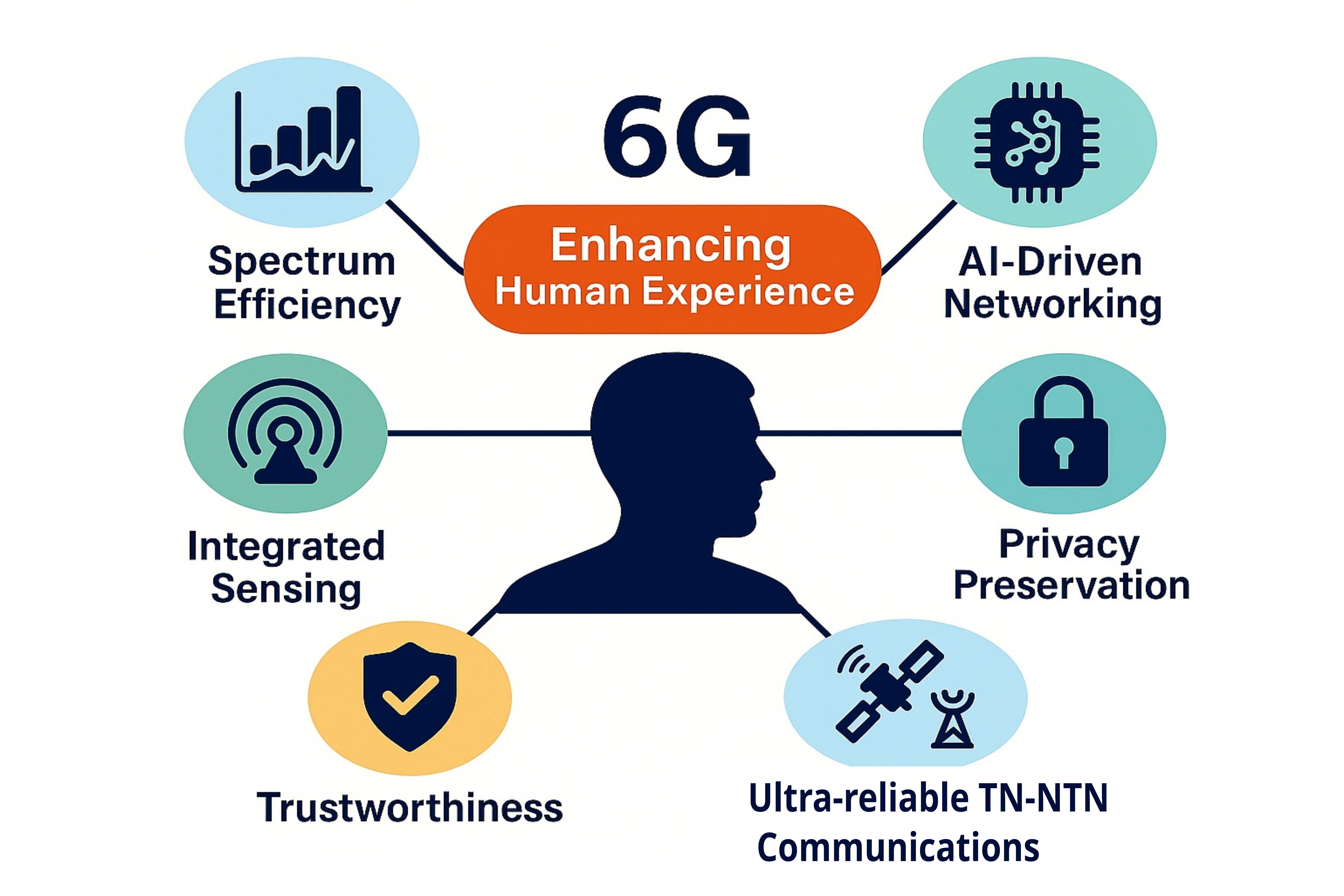}
\caption{\small \textit{Challenges being address in the 6G fortiss research}.}\label{fig:challenges}
\end{figure}

The overall 6G vision integrates the core concept of universal and inclusive connectivity, affordable to all, and of immersive, inclusive communication, cognition, and twinning. The focus is on applications and services that can improve societal well-being based on an advanced intertwining of virtual representations in the digital world of entities in the physical and human worlds. Such virtual representations will serve human needs by offering reasoning capabilities and intelligent connection with sensors and actuators in different forms. Therefore, in terms of data communications, the final 6G architecture needs to address functions that will have to manage not only interconnection between end-user devices but also seamless and intuitive interactions between human users and virtual digital entities, requiring ultra-low latency and extremely high data rates that match human perceptual capabilities~\cite{banafaa20236g}.
This leads to several challenges as represnted in Figure \ref{fig:challenges}.

The possibility to engineer the necessary support requires a paradigm shift that goes beyond the development of advanced hardware. Specifically, XR multi-sensory applications will require tight synchronization, which is not feasible to be achieved simply by improving spectrum diversity. This challenge is further exacerbated by the increased demand for \textit{\ac{ISAC}/\ac{JCAS}}, where wireless signals are used simultaneously for data transmission and environmental awareness, raising concerns regarding interference management, privacy, and regulatory frameworks.
Meeting the demands of a human-centered 6G vision goes far beyond simply scaling connectivity through more devices, higher data rates, or incremental advances in hardware, spectrum, or AI/ML integration. Instead, it requires a fundamental rethinking of immersive, context-aware communication—particularly the seamless integration of terrestrial and \textit{non-terrestrial networks (NTN)}.

Future 6G infrastructures will be multi-layered and distributed across both earth and space, incorporating smart satellite constellations, \textit{\ac{HAP}}, and \textit{\ac{UAV}}. These heterogeneous networks introduce significant challenges, including efficient handovers, signal propagation delays, and the need for seamless end-to-end connectivity. Addressing these issues is crucial to enabling uninterrupted, human-centric services—regardless of user location, activity, or environmental context.

A central challenge in realizing 6G lies in the dynamic orchestration of shared resources across heterogeneous networks. In this context, AI/ML engineering plays a critical role, enabling real-time network optimization and adaptive service delivery. However, the integration of AI/ML also introduces significant concerns.

Foremost among these are energy consumption and data scarcity. Real-time AI-driven operations demand substantial computational resources, which may conflict with 6G’s sustainability goals. Furthermore, issues of explainability, transparency, and bias in AI-driven decision-making present additional risks. As network functions become increasingly automated, it becomes essential to develop trustworthy, interpretable, and fair ML models that support not only effective performance but also accountability and human-aligned outcomes.

Security and privacy preservation become even more critical in the context of 6G. As networks evolve into a dynamic, decentralized edge–cloud continuum, the ability to protect sensitive data across distributed environments is essential. This urgency is amplified by 6G’s anticipated support for emerging technologies such as digital twins, holographic communications, and \textit{\ac{BCI}}, all of which involve the transmission of deeply personal data. These applications raise significant ethical and security concerns, requiring robust, privacy-aware architectures that can ensure data protection without compromising performance or user trust~\cite{hu2024survey}.

Summarising, 6G's success will be measured not only by its technical capabilities but by how well it serves humanity. Beyond supporting vastly increased connectivity demands, it must address fundamental limitations in ways that enhance human experience, promote sustainability, and build trustworthiness through transparent, ethical design. The convergence of spectrum efficiency, AI-driven networking, integrated sensing, security frameworks,  and ultra-reliable terrestrial to non-terrestrial (TN–NTN) communication will define the success of 6G in delivering ubiquitous, intelligent, and sustainable connectivity that enhances human lifes at both the individual and societal level.

\newpage

\section{6G Active Research at fortiss}
\label{6Gresearch}
With the vision of 6G established as a human-centered, intelligent, and sustainable communication paradigm, we now turn to the enabling technological foundations that must evolve to realize this vision. The following subsections explore the architectural, computational, and semantic innovations required to support 6G’s interdisciplinary goals.
\begin{figure}[h!]
\centering
\includegraphics[width=0.6\textwidth]{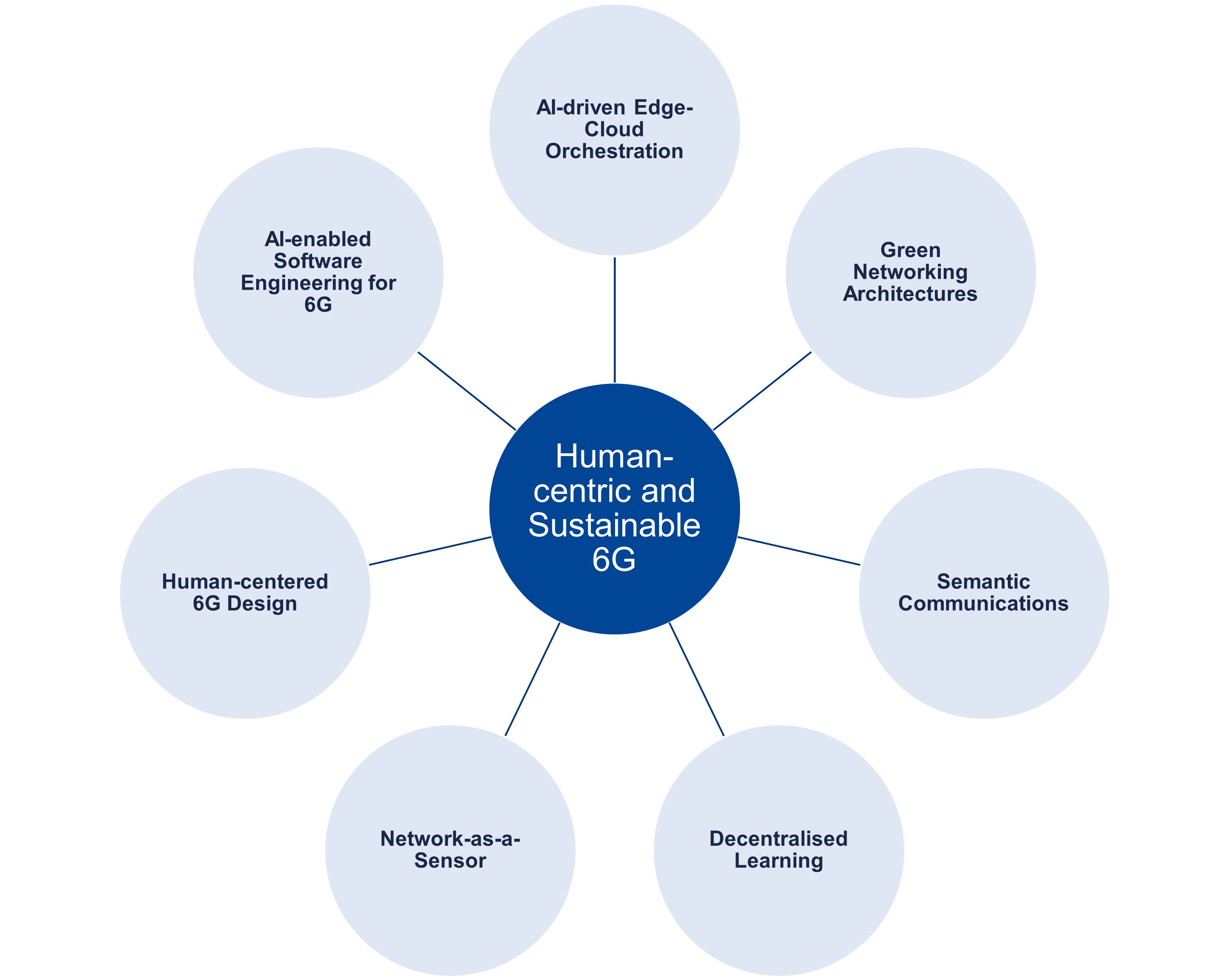}
\caption{\small \textit{Main categories of research being developed at fortiss, envisioning a human-centric and sustainable 6G paradigm}.}\label{fig:areas}
\end{figure}
As a leading research institute, fortiss is at the forefront of efforts to advance sustainable and human-centric 6G technologies and services. Its efforts are structured around three key thematic clusters, each addressing distinct but interrelated aspects of the 6G vision:

\begin{itemize}
    \item \textbf{Software and Systems Engineering:} Developing methods and tools for reliable, secure, and maintainable software systems that can operate in unpredictable environments.
    \item \textbf{AI Engineering:} Creating robust and trustworthy AI technologies for 6G systems that can make safe decisions in uncertain conditions.
   \item \textbf{\ac{IoT} Engineering/Smart Infrastructures:} Developing flexible, software-based infrastructures for 6G that adapt to changing requirements and integrate sensing, computation, and communication capabilities.
\end{itemize}

Current fortiss contributions span several key areas which are represented in Figure \ref{fig:areas}, including \textbf{Edge-Cloud orchestration},
\textbf{green networking}, \textbf{semantic communications}, \textbf{decentralised learning}, \textbf{network-as-a-sensor}, \textbf{human-centered design} and \textbf{AI-enabled software engineering}.

\subsection{AI-driven Edge-Cloud Resource Orchestration}

\textbf{fortiss Focus:} Developing adaptive orchestration strategies for dynamic, privacy-aware resource management across the Edge--Cloud continuum.
\textbf{fortiss Contributes to This Area By:} Designing software-based orchestration mechanisms that jointly manage data, compute, and network resources in multi-tenant, distributed 6G environments.
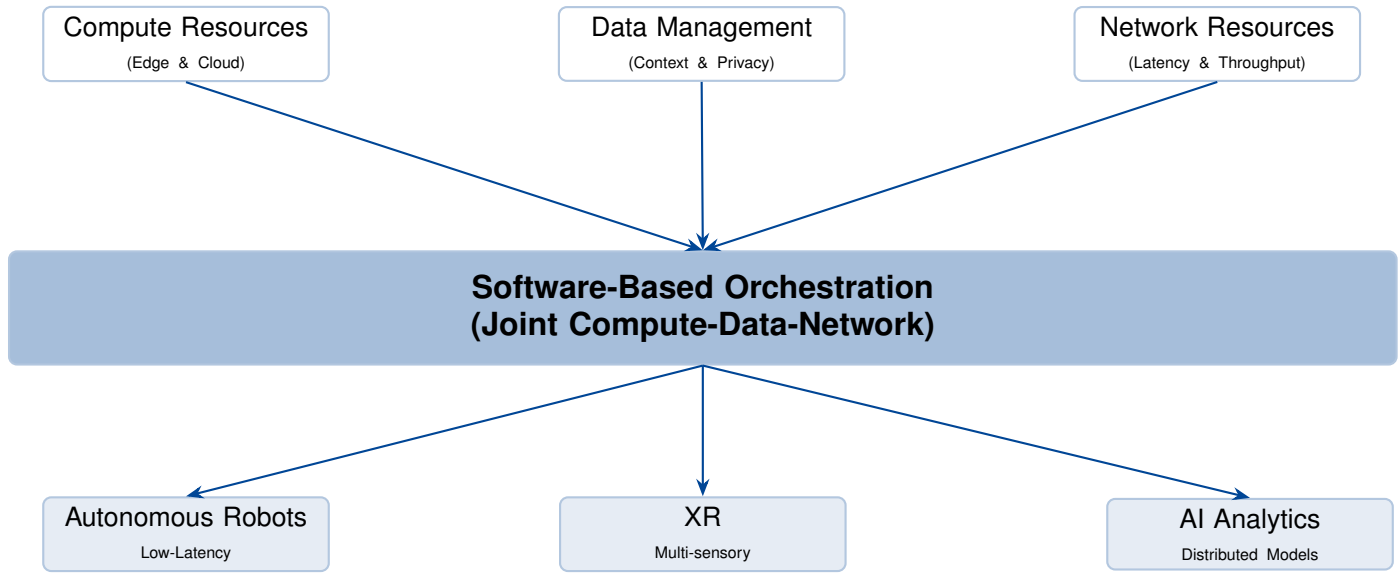
\begin{figure}[h!]
  \centering
  \begin{tikzpicture}[
    xscale=1, yscale=1,
    every node/.style={align=center},
    font=\small,
    block/.style={rectangle, thick, rounded corners=3pt, minimum height=0.9cm, text width=3.5cm},
    layerblock/.style={block, draw=fortissblue!30!white, fill=white!10},
    workloadblock/.style={block, draw=fortissblue!30!white, fill=fortissblue!10},
largeblock/.style={
    block,
    minimum height=1.5cm,     
    text width=\textwidth,           
    font=\large ,              
draw=fortissblue!30!white, fill=fortissblue!35},
 arrow/.style={thick,->,>=Stealth, draw=fortissblue}
  ]
    \node[layerblock] (compute) at (0,3.5) {Compute Resources\\ \tiny (Edge \& Cloud)};
    \node[layerblock, right=3cm of compute] (data) {Data Management\\ \tiny(Context \& Privacy)};
    \node[layerblock, right=3cm of data] (network) {Network Resources\\ \tiny (Latency \& Throughput)};

    \node[largeblock] (orchestration) at (6.8,0){\textbf{Software-Based Orchestration}\\\textbf{(Joint Compute-Data-Network)}};

    \node[workloadblock] (vehicular) at (0,-3) {Autonomous Robots\\ \tiny Low-Latency};
    \node[workloadblock] (xr) at (6.8,-3) {XR \\ \tiny Multi-sensory};
      \node[workloadblock] (analytics) at (14,-3) {AI Analytics\\ \tiny Distributed Models};

    \draw[arrow] (compute.south) -- (orchestration.north);
    \draw[arrow] (data.south) -- (orchestration.north);
    \draw[arrow] (network.south) -- (orchestration.north);

    \draw[arrow] (orchestration.south) -- (vehicular.north);
    \draw[arrow] (orchestration.south) -- (xr.north);
    \draw[arrow] (orchestration.south) -- (analytics.north);

  \end{tikzpicture}
  \caption{\textit{fortiss joint orchestration across compute, data, and network. Infrastructure resources are represented above in light blue; examples of application workloads considered are represented in yelow below the main goal of software-based joint orchestration. Arrows represent direct interdisciplinary input to the specific research goal.}\label{fig:edge-cloud}\cite{sofia2024framework}.}
\end{figure}
Being a next generation communications paradigm, 6G requires flexible and AI-based network management, to adapt to new applications, user behavior, and dense application deployment, while at the same time ensuring data privacy and sovereignty. To fully realize its potential, 6G must address a synergy between shared resources at the \textbf{computational, data, and networking levels}, particularly across the Edge-Cloud continuum. 
This continuum integrates edge computing—bringing computation closer to data sources—with cloud computing’s centralized processing power. The goal is to establish an efficient, scalable framework that meets the diverse requirements of latency-sensitive and high-throughput applications central to 6G, such as IoT, autonomous systems, and multi-sensory XR. For this and as represented in Figure \ref{fig:edge-cloud}, several topics are being worked upon, as explained next.

 Dynamic deployment across the Edge–Cloud continuum will be a cornerstone of 6G, ensuring that compute, storage, and networking resources are allocated in real time to meet application-specific needs. A key challenge in this context—one fortiss is actively addressing—is the \textbf{joint orchestration of computational, network and data} resources in multi-tenant, distributed environments. This requires adaptive strategies that account for varying application profiles. For instance, autonomous vehicles demand ultra-low-latency, real-time processing at the edge, while AI-driven analytics often benefit from federated or decentralized models spanning edge and cloud.

The ability to manage this interplay between edge and cloud resources will be essential for creating a network that can dynamically adapt to varying workload demands and environmental conditions. 

A significant contribution being developed by fortiss in this context relates with the development of software-based orchestration mechanisms that rely on a data-compute-network orchestration approach~\cite{sofia2024framework}\footnote{\url{https://zenodo.org/communities/he-codeco/records?q=&l=list&p=1&s=10&sort=newest}}.

\subsection{Green Networking Architectures}
\textbf{fortiss Focus:} Enabling energy-efficient, cross-layer orchestration across the Edge--Cloud continuum to meet the growing demands of data-intensive and AI-enhanced 6G services.

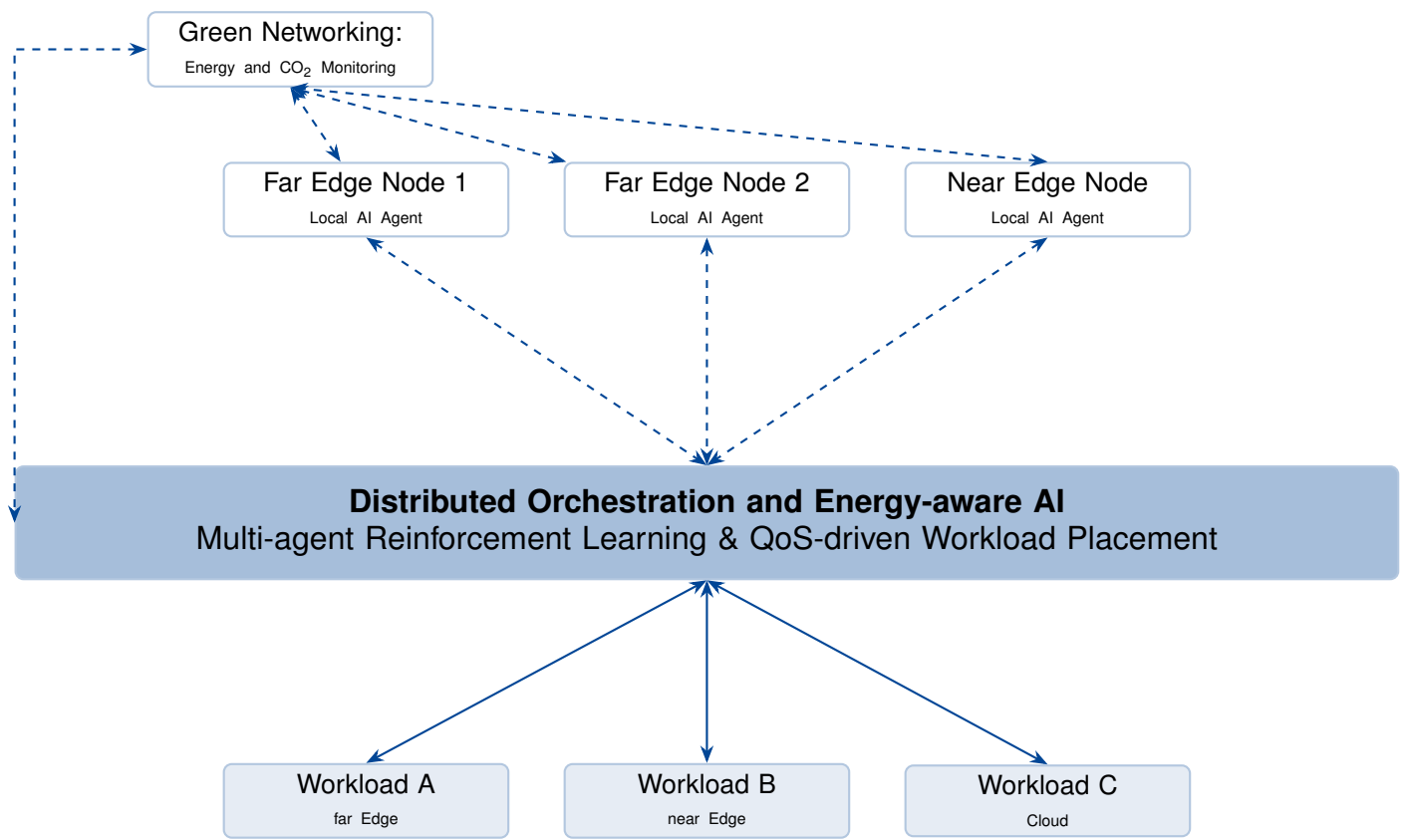
\begin{figure}[h!]
  \centering
  \begin{tikzpicture}[
    xscale=1, yscale=1,
    every node/.style={align=center},
    font=\small,
      block/.style={rectangle, thick, rounded corners=3pt, minimum height=0.9cm, text width=3.5cm},
    layerblock/.style={block, draw=fortissblue!30!white, fill=white!10},
    workloadblock/.style={block, draw=fortissblue!30!white, fill=fortissblue!10},
    largeblock/.style={
    block,
    minimum height=1.5cm,     
    text width=\textwidth,           
    font=\large ,              
  draw=fortissblue!30!white,fill=fortissblue!35},
  arrow/.style={thick,->,>=Stealth, draw=fortissblue},
    arrow/.style={thick,<->,>=Stealth, draw=fortissblue},
    dashedarrow/.style={arrow, dashed}
  ]
  \node[layerblock] (edge1) at (0,0) {Far Edge Node 1\\\tiny Local AI Agent};
  \node[layerblock] (edge2) at (4.5,0) {Far Edge Node 2\\ \tiny Local AI Agent};
  \node[layerblock] (edge3) at (9,0) {Near Edge Node\\ \tiny Local AI Agent};

  \node[workloadblock] (workload1) at (0,-8) {Workload A \\ \tiny far Edge};
  \node[workloadblock] (workload2) at (4.5,-8) {Workload B \\ \tiny near Edge};
  \node[workloadblock] (workload3) at (9,-8) {Workload C \\ \tiny Cloud};

  \node[largeblock] (orchestration) at (4.5,-4.3) {
    \textbf{Distributed Orchestration and Energy-aware AI}\\
    Multi-agent Reinforcement Learning \& QoS-driven Workload Placement
  };

  \node[layerblock] (monitor) at (-1,2) {
    Green Networking:\\ \tiny Energy and CO\textsubscript{2} Monitoring
  };

  \draw[dashedarrow] (edge1.south) -- (orchestration.north);
  \draw[dashedarrow] (edge2.south) --  (orchestration.north);
  \draw[dashedarrow] (edge3.south) -- (orchestration.north);

  \draw[arrow] (orchestration.south) --  (workload1.north);
  \draw[arrow] (orchestration.south) -- (workload2.north);
  \draw[arrow] (orchestration.south) -- (workload3.north);

  \draw[dashedarrow] (monitor.west) -| (orchestration.west);

  \draw[dashedarrow] (edge1) -- (monitor.south);
  \draw[dashedarrow] (edge2) -- (monitor.south);
  \draw[dashedarrow] (edge3.north) -- (monitor.south);
  
  \end{tikzpicture}
  \caption{
  \textit{  Energy-aware orchestration and workload deployment across the Edge–Cloud continuum.
    fortiss contributes with multi-agent AI, reinforcement learning, and QoS models to optimize resource allocation under energy and carbon constraints. Dashed arrows represent control data; stealth bi-directional arrows represent service/data plane data.}
  }
  \label{greennet}
\end{figure}

\textbf{fortiss Contributes to This Area By:} Developing energy-aware QoS models and orchestration strategies that optimize resource allocation based on workload, network conditions, and carbon footprint.
Another key challenge that 6G must address is the design of an \textbf{energy-efficient, cross-layer architecture} for the Edge–Cloud continuum~\cite{c2024shaping}. A representation on how fortiss is proposing to address this challenge is provided in Figure~\ref{greennet}

As data generation and transmission grow exponentially driven largely by the proliferation of IoT devices, the energy consumption of the entire infrastructure, including both hardware and software components across devices and networks, has become a critical concern. Integrating AI/ML technologies and workloads increases complexity in 6G networks.

Energy efficiency will depend not only on individual assets (e.g., devices, data centers) but also on the computational and network resources needed to maintain resilience and reliability. Consequently, energy efficiency must be a core principle in the design of flexible Edge–Cloud orchestration strategies. 

For instance, Edge computing nodes can help reduce data transmission costs and energy consumption by processing data locally, while Cloud data centers can be optimized for energy efficiency using cross-layer AI-based workload distribution strategies, e.g., \textit{\ac{MARL}}. In this context, cross-layer design refers to the intelligent optimization of energy usage across edge and cloud environments, where the system dynamically determines where to process, store, and transmit data based on real-time network conditions application workload demands, and overall energy consumption, or CO2 footprinting.
Figure~\ref{greennet} shows a high-level representation of an Edge-Cloud infrastructure, where multiple Edge nodes (Far and Near Edge) are equipped with local AI agents and collaborate with a distributed orchestration layer. This orchestration layer, powered, for instance, by MARL and energy-aware QoS models, dynamically manages workload placement and resource allocation across Edge-Cloud. Control data, represented by dashed arrows, flows between Edge nodes, the orchestration layer, and a green networking monitor enable intelligent, real-time adaptation based on energy consumption and CO₂ emission metrics. Application workloads are then placed across Edge-Cloud based on decisions developed in the AI-driven, energy-aware orchestration layer (stealth arrows).

This concept embodies the principle of \textit{cross-layer optimization}: Edge devices can process data locally to minimize transmission energy, while Cloud-scale data centers optimize large-scale processing and storage using AI-based strategies. The energy and CO$_2$ monitoring block informs these decisions by feeding real-time environmental metrics back into orchestration and workload deployment logic.

In this context, fortiss’s contribution focuses on designing orchestration strategies that are \textbf{energy-efficient by design}, integrating both performance and sustainability constraints. These strategies rely on flexible, scalable coordination mechanisms that optimize computing and network resources holistically, across all system layers.
Ongoing contributions in this context relate with the definition and application of energy QoS models, and on resource orchestration that is energy-efficient by design \cite{sofia2025green}. 

\subsection{Semantic Communications}
\textbf{fortiss Focus:} Advancing context-aware semantic communication systems that prioritize intent and meaning over raw data to enhance efficiency in IoT–Edge–Cloud ecosystems.


\textbf{fortiss Contributes to This Area By:} Developing semantic communication architectures tailored for \textit{\ac{IIoT}}, enabling intelligent data prioritization, reduced redundancy, and improved QoS/QoE in real-time applications.
Semantic communication architectures will play a critical role in enabling efficient information exchange in 6G, especially across the IoT–Edge–Cloud continuum. As IoT devices and sensors continue to generate vast volumes of data, there is a growing need to move beyond conventional bit-level transmission. Semantic communication focuses on conveying the meaning or intent of information, rather than simply transmitting raw data bits.

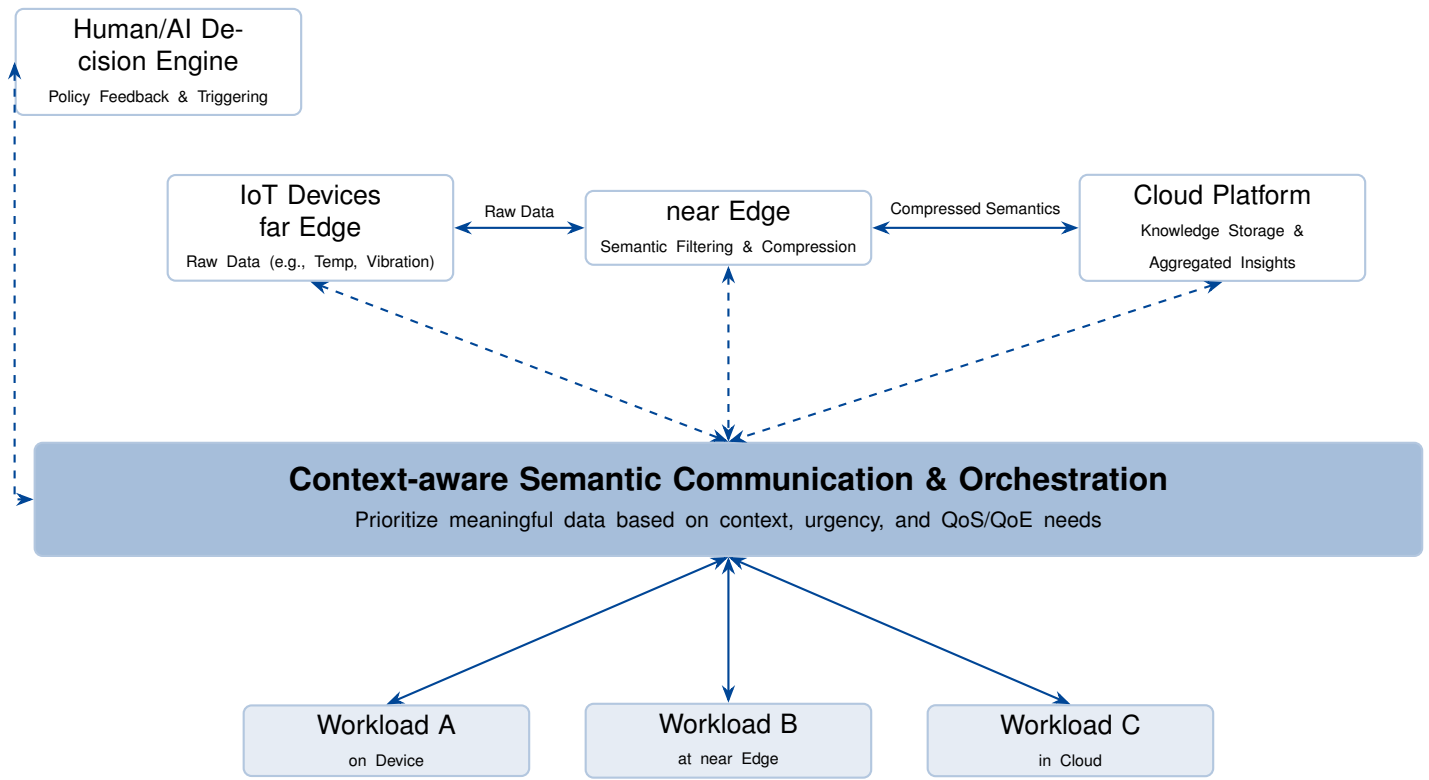
\begin{figure}[h!]
  \centering
  \begin{tikzpicture}[
    xscale=1, yscale=1,
    every node/.style={align=center},
    font=\small,
      block/.style={rectangle, thick, rounded corners=3pt, minimum height=0.9cm, text width=3.5cm},
    layerblock/.style={block, draw=fortissblue!30!white, fill=white!10},
    workloadblock/.style={block, draw=fortissblue!30!white, fill=fortissblue!10},
    largeblock/.style={
    block,
    minimum height=1.5cm,     
    text width=\textwidth,           
    font=\large ,              
  draw=fortissblue!30!white,fill=fortissblue!35},
  arrow/.style={thick,->,>=Stealth, draw=fortissblue},
    arrow/.style={thick,<->,>=Stealth, draw=fortissblue},
    dashedarrow/.style={arrow, dashed}
  ]
  \node[layerblock] (edge1) at (-1,0) {IoT Devices\\far Edge\\\tiny Raw Data (e.g., Temp, Vibration)};
  \node[layerblock] (filter) at (4.5,0) {near Edge\\\tiny Semantic Filtering \& Compression};
  \node[layerblock] (cloud) at (11,0) {Cloud Platform\\\tiny Knowledge Storage \& Aggregated Insights};

  \node[workloadblock] (workload1) at (0,-6.8) {Workload A\\ \tiny on Device};
  \node[workloadblock] (workload2) at (4.5,-6.8) {Workload B\\ \tiny at near Edge};
  \node[workloadblock] (workload3) at (9,-6.8) {Workload C\\ \tiny in Cloud};

  \node[largeblock] (orchestration) at (4.5,-3.6) {
    \textbf{Context-aware Semantic Communication \& Orchestration}\\
    \scriptsize Prioritize meaningful data based on context, urgency, and QoS/QoE needs
  };

  \node[layerblock] (monitor) at (-3,2.2) {Human/AI Decision Engine\\\tiny Policy Feedback \& Triggering};

  \draw[arrow] (edge1.east) -- node[above] {\tiny Raw Data} (filter.west);
  \draw[arrow] (filter.east) -- node[above] {\tiny Compressed Semantics} (cloud.west);

  \draw[dashedarrow] (edge1.south) --  (orchestration.north);
  \draw[dashedarrow] (filter.south) -- (orchestration.north);
  \draw[dashedarrow] (cloud.south) -- (orchestration.north);

  \draw[arrow] (orchestration.south) -- (workload1.north);
  \draw[arrow] (orchestration.south) -- (workload2.north);
  \draw[arrow] (orchestration.south) -- (workload3.north);

  \draw[dashedarrow] (monitor.west) -- ++(0,-1) |- (orchestration.west);


  \end{tikzpicture}
  \caption{
    \textit{Semantic communication architecture for the IoT–Edge–Cloud continuum. Data is semantically filtered at the edge before transmission. fortiss focuses on context-aware orchestration to reduce redundancy and improve QoS/QoE, with decision loops from AI/Human systems feeding back into data prioritization. Dashed arrows reorese}
  }
  \label{fig:semcomiot}
\end{figure}

This shift allows for more context-aware and purpose-driven communication, significantly reducing data redundancy and optimizing network resource usage. For instance, instead of transmitting complete sensor datasets, a semantic communication system as represented in Figure \ref{fig:semcomiot} may send only the insights necessary to support a decision or trigger an action. This not only reduces network congestion but also enhances system responsiveness and overall energy efficiency.

At fortiss, the research in semantic communications focuses on the analysis and integration of semantic communication architectures into the evolving Edge-Cloud continuum and IoT ecosystems\footnote{\url{https://www.fortiss.org/en/research/projects/detail/semcomiiot}}. 
The focus is on analyzing how communication systems can become more context-aware, enabling smarter decisions about what data to transmit, when, and how to prioritize it. Context-aware communication allows the system to assess the relevance of information in real time, improving efficiency by transmitting only what is most meaningful for the task at hand.

For example, in a smart city scenario, traffic monitoring systems could transmit high-level insights—such as congestion alerts or accident reports—instead of streaming continuous raw traffic data. This targeted communication reduces bandwidth usage, lowers energy consumption, and supports faster, more informed decision-making.

A first aspect being addressed relates with the development of \textbf{context-aware IoT and communication systems}. The goal is to develop systems capable of understanding not only the data itself but also the intent behind its transmission. This enables more efficient use of network resources by allowing data to be prioritized based on its importance, urgency, and contextual relevance to a specific user or application. Such an approach can significantly enhance Quality of Service (QoS) for both real-time and non-real-time applications, ensuring that critical information is delivered promptly while minimizing unnecessary data transmission.

A second key aspect under development is the capability to reduce data redundancy. By emphasizing the meaning of the information rather than the raw data itself, semantic communication systems aim to transmit only the most relevant portions of data—eliminating the need to send excessive or redundant information. This is particularly relevant to dense IoT environments, where millions of connected devices continuously generate large volumes of data. Traditional communication systems typically transmit all collected data, which can lead to network congestion and inefficiencies. In contrast, semantic systems can intelligently filter and prioritize transmissions based on contextual relevance, enabling more efficient use of bandwidth and reducing the processing burden on network infrastructure.

For example, in an industrial IoT scenario, machinery sensors can communicate the current status of the system, but rather than sending raw data constantly, the system might only transmit information when anomalies or threshold limits are exceeded, drastically reducing the communication overhead and improving overall network efficiency.

From a practical standpoint, the semantic communications research at fortiss focuses on the development of semantic communication architectures tailored to support IIoT services.

\subsection{Decentralized Learning for a Flexible Control Plane}
\textbf{fortiss Focus:} Advancing decentralized, AI-powered cognitive control planes to enable adaptive, autonomous decision-making across the Edge--Cloud continuum in 6G networks.

\textbf{fortiss Contributes to This Area By:} Developing scalable AI-as-a-Service frameworks and real-time, distributed learning strategies—including reinforcement learning and split inference—to support intelligent orchestration in dynamic, resource-constrained 6G environments.

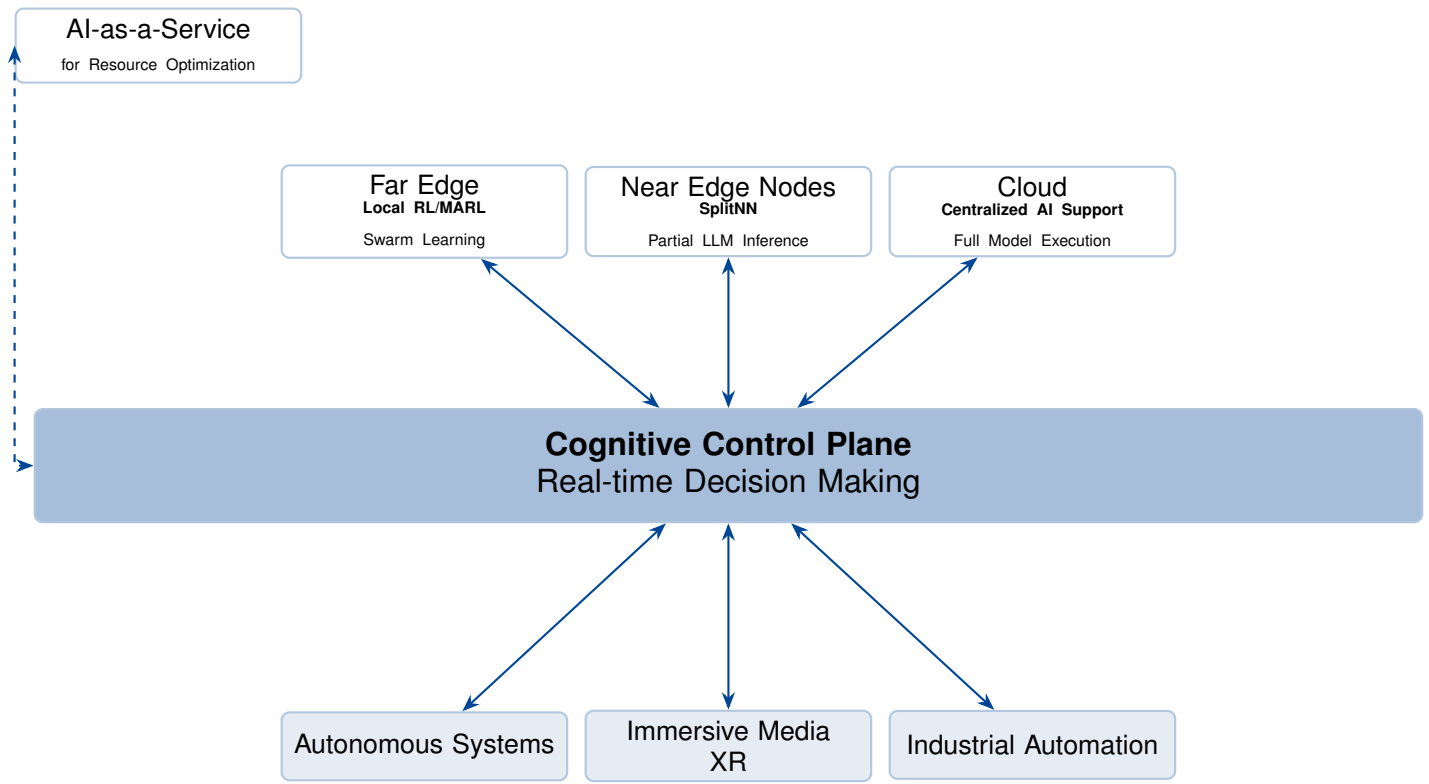
\begin{figure}[h!]
  \centering
  \begin{tikzpicture}[
   xscale=1, yscale=1,
    every node/.style={align=center},
    font=\small,
      block/.style={rectangle, thick, rounded corners=3pt, minimum height=0.9cm, text width=3.5cm},
    layerblock/.style={block, draw=fortissblue!30!white, fill=white!10},
    workloadblock/.style={block, draw=fortissblue!30!white, fill=fortissblue!10},
    largeblock/.style={
    block,
    minimum height=1.5cm,     
    text width=\textwidth,           
    font=\large ,              
  draw=fortissblue!30!white,fill=fortissblue!35},
  arrow/.style={thick,->,>=Stealth, draw=fortissblue},
    arrow/.style={thick,<->,>=Stealth, draw=fortissblue},
    dashedarrow/.style={arrow, dashed}
  ]
    \node[layerblock] (edge) at (-4,0) {Far Edge \\ \tiny \textbf{Local RL/MARL} \\ \tiny Swarm Learning};
    \node[layerblock] (nearEdge) at (0,0) {Near Edge Nodes\\ \tiny \textbf{SplitNN} \\ \tiny  Partial LLM Inference};
    \node[layerblock] (cloud) at (4,0) {Cloud\\ \tiny  \textbf{Centralized AI Support}\\ \tiny Full Model Execution};

    \node[largeblock, below=2cm of nearEdge] (control) {\textbf{Cognitive Control Plane}\\ Real-time Decision Making};

    \node[workloadblock, below=6cm of edge] (auto) {Autonomous Systems};
    \node[workloadblock, below=6cm of nearEdge] (xr) {Immersive Media \\ XR};
    \node[workloadblock, below=6cm of cloud] (industry) {Industrial Automation};

    \draw[arrow] (edge) -- (control);
    \draw[arrow] (nearEdge) -- (control);
    \draw[arrow] (cloud) -- (control);

    \draw[arrow] (control) -- (auto);
    \draw[arrow] (control) -- (xr);
    \draw[arrow] (control) -- (industry);

    \node[layerblock] (aiaas) at (-7.5,2.2) {AI-as-a-Service\\\tiny for Resource Optimization};
    
    \draw[dashedarrow] (aiaas.west) -- ++(0,-1) |- (control.west);

  \end{tikzpicture}
  \caption{\textit{Decentralized Cognitive Control Plane with AI-as-a-Service and distributed learning across Edge–Cloud 6G systems. The figure illustrates a decentralized cognitive control plane that coordinates AI-driven decision-making across Far Edge, Near Edge, and Cloud layers. Edge layers support local learning (e.g., RL, MARL, SplitNN), while the Cloud enables full model execution. AI-as-a-Service modules can trigger optimization or adaptation strategies across layers. The control plane integrates real-time data flows and semantic prioritization for key applications such as immersive media (XR), industrial automation, and autonomous systems. The dashed arrows represent the control data; the solid arrows represent the data plane.
}\label{fig:decentralisedcontrolplane}}
\end{figure}

The cognitive control plane will be a foundational component of 6G networks, especially within the decentralized Edge–Cloud continuum. Its role is to dynamically manage network operations based on real-time data, enabling the system to learn from its environment and make intelligent, context-aware decisions. As 6G integrates a diverse range of technologies across an Edge-Cloud continuum with elements both on earth and space, the control plane must operate autonomously, as represented in Figure \ref{fig:decentralisedcontrolplane}.

This autonomy is essential for coordinating resources, optimizing traffic flow, and maintaining security without relying on centralized control. Such a decentralized approach will enhance the network’s resilience, scalability, and adaptability, allowing it to respond to local conditions and user-specific demands in real time. The cognitive control plane will also be critical for enabling a wide array of advanced 6G applications, including autonomous vehicles, immersive media, and industrial automation, all of which require intelligent, responsive, and low-latency network behavior.

Current AI/ML centralized approaches are not sufficient to support the required 6G flexibility, as it will collide with scalability and fault tolerance requirements of 6G.

An emerging trend is the provision of \textit{AI-as-a-Service} by network operators. In this model, AI-driven capabilities are offered as on-demand services to support the management of increasingly complex network environments~\cite{lins2021artificial}. 

AI-as-a-Service address critical areas such as network resource optimization and configuration automation, enabling more agile, efficient, and responsive systems. For instance, AI can dynamically allocate bandwidth, predict traffic surges, or automate network configurations based on real-time analysis, significantly reducing manual intervention and improving operational efficiency.
The current research being developed aims to address AI engineering across 6G mobile environments, considering its deployment from the far Edge to the Cloud.

The concept of AI-as-a-Service offers significant potential to enhance the agility, scalability, and functionality of 6G networks. By delivering AI capabilities as modular, on-demand services, network operators can support a broad spectrum of applications for diverse stakeholders—from enterprises to individual users. These services might include predictive analytics for network optimization, automated fault detection and resolution for improved service reliability, or AI-driven user interfaces for personalized network configuration.

Hence, AI-as-a-Service not only democratizes access to advanced AI tools but also fosters innovation by enabling smaller organizations and developers to leverage cutting-edge technologies without the need for large-scale infrastructure. In doing so, it creates a more inclusive and dynamic 6G ecosystem, where intelligence is both accessible and adaptable to specific user needs.

A trend in this context focuses on addressing up to which point can \textbf{decentralized or hierarchical AI/ML} approaches be helpful. Specifically, Swarm Learning~\cite{wang2024distributed}, SplitNNs~\cite{gao2020end} and Gossip Learning are relevant in the context of knowledge exchange across the Edge-Cloud.

A related trend involves the adaptation of \textit{\ac{RL}} and \textit{\ac{MARL}} to the Edge–Cloud continuum, with a focus on training and adapting personalized models to specific environments while ensuring scalability in multi-tenant scenarios. A key challenge here is moving beyond traditional offline, centralized training approaches. Instead, there is growing emphasis on enabling inline (real-time) training at the far edge—where data is generated—so that learning can occur in dynamic, resource-constrained environments.

To support this, current research explores the design of a decentralized learning and decision-making layer integrated with the Edge–Cloud control plane. This layer facilitates partial, localized decision-making and continuous adaptation of the system, allowing edge clusters to maintain operational isolation while still sharing sufficient resources to support a self-organizing, scalable ecosystem. Such an approach is critical to achieving autonomous, context-aware behavior across distributed 6G environments.

Central to this exploration is the potential role of \textbf{large foundational models}, which have demonstrated remarkable capabilities in natural language processing, semantic understanding, and adaptive decision-making, to assist the flexible orchestration of 6G~\cite{xiao2024llm}. Initial research in this context relates with network optimization aspects, in particular concerning radio resource sharing. 

However, there are critical challenges when considering the deployment of large foundational models at the edge (currently, near Edge), in particular related with training and inference due to the limited environments at the Edge~\cite{lin2023pushing}. Different techniques are relevant to be considered in this context, e.g., split learning, quantization.

\subsection{Network as a Sensor}
\textbf{fortiss Focus:} Exploring the fusion of communication and sensing to enable intelligent, perception-aware 6G networks capable of environmental awareness and precise localization.

\textbf{fortiss Contributes to This Area By:} Advancing research in ISAC/JCAS and privacy-aware 6G localization, including \textit{\ac{SLAM}} applications, to support autonomous systems, XR, and secure context-driven services.

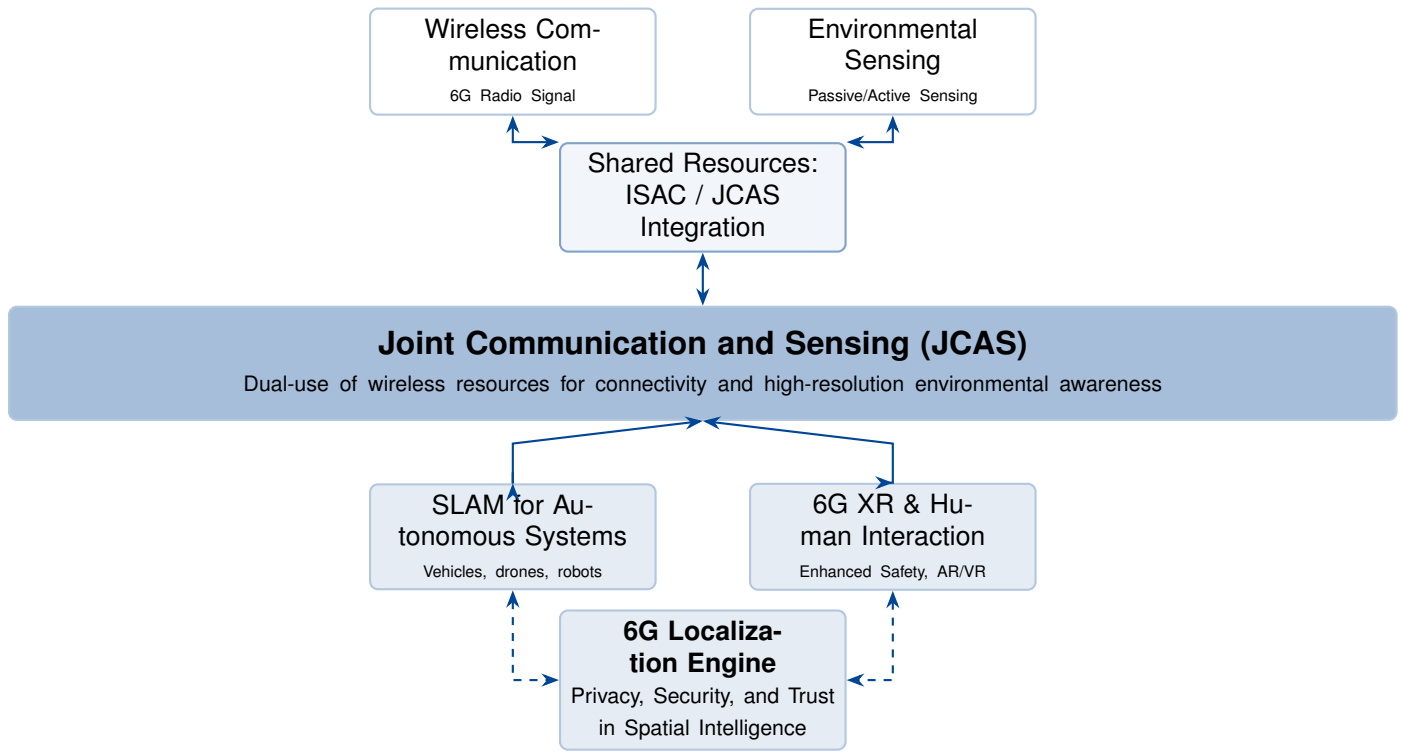
\begin{figure}[h!]
  \centering
  \begin{tikzpicture}[
    xscale=1, yscale=1,
    every node/.style={align=center},
    font=\small,
      block/.style={rectangle, thick, rounded corners=3pt, minimum height=0.9cm, text width=3.5cm},
    layerblock/.style={block, draw=fortissblue!30!white, fill=white!10},
    workloadblock/.style={block, draw=fortissblue!30!white, fill=fortissblue!10},
    largeblock/.style={
    block,
    minimum height=1.5cm,     
    text width=\textwidth,           
    font=\large ,              
  draw=fortissblue!30!white,fill=fortissblue!35},
  arrow/.style={thick,->,>=Stealth, draw=fortissblue},
    arrow/.style={thick,<->,>=Stealth, draw=fortissblue},
    dashedarrow/.style={arrow, dashed}
  ]

  \node[layerblock] (comm) at (0,0) {Wireless Communication\\\tiny 6G Radio Signal};
  \node[layerblock] (sense) at (5,0) {Environmental Sensing\\ \tiny Passive/Active Sensing};
  \node[block, draw=fortissblue!50, fill=fortissblue!5] (shared) at (2.5,-1.8) {Shared Resources:\\ISAC / JCAS Integration};

  \node[largeblock] (fusion) at (2.5,-4) {
    \textbf{Joint Communication and Sensing (JCAS)}\\
    \scriptsize Dual-use of wireless resources for connectivity and high-resolution environmental awareness
  };

  \node[workloadblock] (slam) at (0,-6.3) {SLAM for Autonomous Systems\\\tiny Vehicles, drones, robots};
  \node[workloadblock] (xr) at (5,-6.3) {6G XR \& Human Interaction\\\tiny Enhanced Safety, AR/VR};

  \node[workloadblock] (loc) at (2.5,-8.2) {
    \textbf{6G Localization Engine}\\
    \scriptsize Privacy, Security, and Trust in Spatial Intelligence
  };

  \draw[arrow] (comm.south) -- ++(0,-0.3) |- (shared.north west);
  \draw[arrow] (sense.south) -- ++(0,-0.3) |- (shared.north east);
  \draw[arrow] (shared.south) -- (fusion.north);

  \draw[arrow] (fusion.south) -- ++(-2.5,-0.3) |- (slam.north);
  \draw[arrow] (fusion.south) -- ++(2.5,-0.3) |- (xr.north);

  \draw[dashedarrow] (slam.south) -- ++(0,-0.3) |- (loc.west);
  \draw[dashedarrow] (xr.south) -- ++(0,-0.3) |- (loc.east);

  \end{tikzpicture}
  \caption{
    \textit{fortiss research in JCAS, , where wireless signals are used not only for connectivity but also for environmental awareness. Communication and sensing components (top boxes) feed into a shared JCAS integration layer, which enables advanced applications such as SLAM for autonomous systems and extended reality (XR). These applications, in turn, support a privacy-aware 6G Localization Engine. Light boxes represent system components such as communication, sensing, and shared resource integration. Darker boxes represent target applications or joint functionalities. Solid arrows show the main data and function flow between components, while dashed arrows represent semantic, contextual, or inferred data relationships contributing to localization.}
  }
  \label{fig:netsensor}
\end{figure}

Another game-changing development of 6G technology is to consider the network infrastructure as an advanced sensing infrastructure and enable it to perceive its overall environment and context without requiring active sensors and actuators. Key components for this development are represented in Figure \ref{fig:netsensor}. This capability is seen as a natural evolution of existing technologies like radar and IoT sensors. By leveraging radio frequency reflections, 6G networks can detect objects, their properties, and movements, creating a "digital twin'' of the physical world.

Such a 6G network sensor is expected to revolutionize communication networks by extending human awareness far beyond physical limitations and enabling a wide array of innovative services. 

Two relevant areas being addressed in fortiss comprise ISAC/JCAS and 6G localization.

ISAC/JCAS integrates wireless communication with environmental sensing, enabling networks to detect and map surroundings without active sensors. Utilizing higher frequencies, broader bandwidths, and advanced antennas, ISAC/JCAS transforms radio signals into tools for precise object detection and mapping. This dual capability enables applications like enhanced safety, autonomous navigation, and advanced human-machine interactions. By sharing resources between communication and sensing, JCAS optimizes performance, merging connectivity with environmental awareness to revolutionize wireless networks.

A promising application of 6G localization is the \textit{Simultaneous Localization and Mapping (SLAM)} problem, which could enhance navigation for autonomous vehicles and drones, and support advanced cross-reality (XR) experiences. Additionally, 6G radar and communication systems will work together to provide a rich, accurate virtual map of the environment, transforming industries reliant on precise location data.

Our foundational research challenges in 6G localization address privacy, security, and trust, calling for interdisciplinary efforts to unlock the full potential of 6G localization.



\subsection{Human-centered 6G design}
\textbf{fortiss Focus:} Designing human-centered 6G systems that emphasize transparency, explainability, and collaborative intelligence in mission-critical environments.

\textbf{fortiss Contributes to This Area By:} Developing intent-aware, human-in-the-loop decision support mechanisms that integrate symbolic reasoning and causal inference to build trust, mitigate over-reliance, and ensure safe, ethical AI deployment in 6G networks.

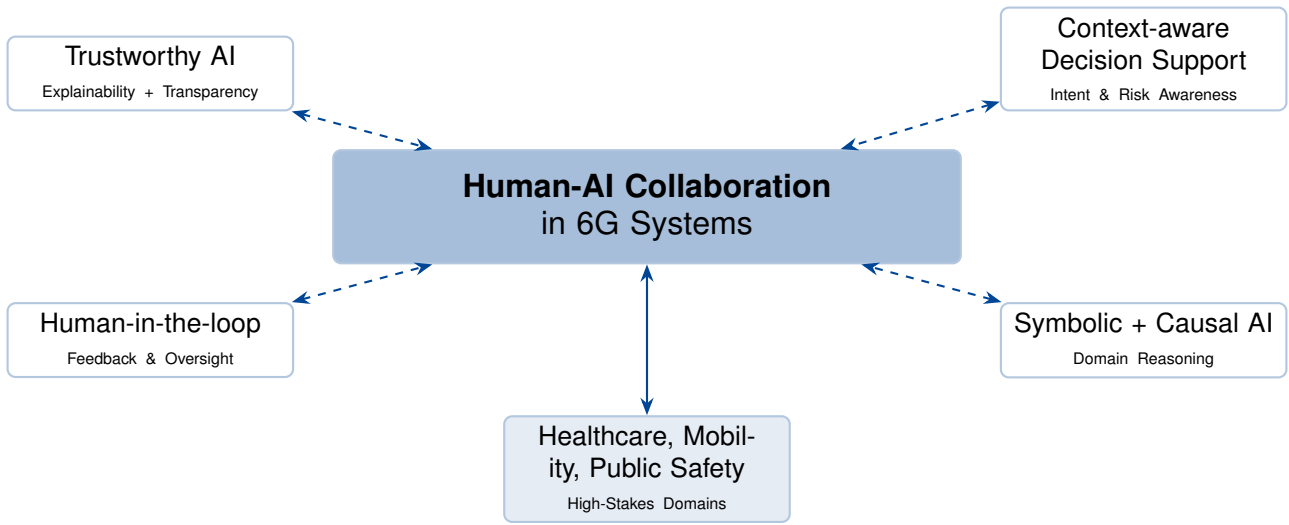
\begin{figure}[h!]
  \centering
  \begin{tikzpicture}[
    xscale=1, yscale=1,
    every node/.style={align=center},
    font=\small,
      block/.style={rectangle, thick, rounded corners=3pt, minimum height=0.9cm, text width=3.5cm},
    layerblock/.style={block, draw=fortissblue!30!white, fill=white!10},
    workloadblock/.style={block, draw=fortissblue!30!white, fill=fortissblue!10},
    largeblock/.style={
    block,
    minimum height=1.5cm,     
    text width=8cm,           
    font=\large ,              
  draw=fortissblue!30!white,fill=fortissblue!35},
  arrow/.style={thick,->,>=Stealth, draw=fortissblue},
    arrow/.style={thick,<->,>=Stealth, draw=fortissblue},
    dashedarrow/.style={arrow, dashed}
  ]
    \node[largeblock] (humanAI) at (0,0) {\textbf{Human-AI Collaboration}\\in 6G Systems};

    \node[layerblock, above left=0.5cm and 0.5cm of humanAI] (trust) {Trustworthy AI\\\tiny Explainability + Transparency};
    \node[layerblock, above right=0.5cm and 0.5cm of humanAI] (context) {Context-aware Decision Support\\ \tiny Intent \& Risk Awareness};

    \node[layerblock, below left=0.5cm and 0.5cm of humanAI] (hitl) {Human-in-the-loop\\ \tiny Feedback \& Oversight};
    \node[layerblock, below right=0.5cm and 0.5cm of humanAI] (symbolic) {Symbolic + Causal AI\\ \tiny Domain Reasoning};

    \node[workloadblock, below=2cm of humanAI] (domains) {Healthcare, Mobility, Public Safety\\\tiny High-Stakes Domains};

    \draw[dashedarrow] (trust) -- (humanAI);
    \draw[dashedarrow] (context) -- (humanAI);
    \draw[dashedarrow] (hitl) -- (humanAI);
    \draw[dashedarrow] (symbolic) -- (humanAI);
    \draw[arrow] (humanAI) -- (domains);
  \end{tikzpicture}
  \caption{\textit{Human-centered design in 6G: enabling trust, context-awareness, and adaptive intelligence for relevant competitiveness domains. Light boxes represent functional modules contributing to the system's cognitive and ethical capacity. Dashed arrows indicate cognitive and contextual input into the collaboration core, while the solid arrow represents the downstream influence on applications across different domains.}}\label{fig:humancentdesign}
\end{figure}

Human-centered principles are emerging as foundational for 6G systems that are transparent, explainable, and capable of fostering meaningful human-AI collaboration. Beyond autonomous network optimization, 6G should enable collaborative intelligence in next-generation communication ecosystems, enhancing human decisions-making, fostering trust, and operating transparently in high-stakes and dynamic contexts such as Healthcare, Mobility, and Public Safety, as represented in Figure \ref{fig:humancentdesign}.

Explainability and transparency are crucial as AI increasingly governs network functions and control such as adaptive resource management, predictive service provisioning, and integrated sensing. The objective is to develop systems that align with user's mental models, expectations, and domain-specific reasoning processes. Some of our researches demonstrate the importance of calibrating trust appropriately, enabling users to meaningfully engage with AI explanations rather than blindly following automated decisions. This is essential for user trust-building and regulatory compliance and safe deployment in high-stakes environments.

Effective human-AI collaboration in 6G requires adaptive decision support mechanisms that move beyond black-box optimization toward intent-aware and context-sensitive models by being aware of the risk of over-reliance. These models must account for human goals, particularly in complex scenarios like emergency response, where decisions must balance technical efficiency with fairness, risk, and well-being. System design must address the risk of over-reliance, a form of maladaptive trust where users defer excessively to AI systems. Research can integrate symbolic reasoning and causal inference into real-time decision pipelines to better reflect human judgment. 

\textit{Human-in-the-loop (HitL)} mechanisms are transforming 6G system design, balancing AI autonomy with human oversight through human expertise, contextual understanding, and real-time feedback. This approach is essential as AI governs mission-critical 6G functions including dynamic spectrum allocation, edge-cloud orchestration, semantic communication, and predictive quality-of-service management. This approach moves beyond reactive monitoring to proactive co-decision, particularly vital given 6G's extreme performance demands and unprecedented complexity. In decentralized 6G environments, research focuses on orchestrating human input across multiple abstraction levels while maintaining privacy and security. This approach redefines 6G architecture by creating systems that learn with, from, and for humans—shifting from automation-centric to human-centered design where AI augments rather than replaces human insight. 

As 6G evolves beyond technical advancements, human-centered approaches becomes one core research priority aligning with human values. 

\subsection{AI-enabled Software Engineering for 6G}

\textbf{fortiss Focus:} Investigating how AI-assisted software engineering can accelerate 6G development by supporting developers in navigating novel and complex system architectures.

\begin{figure}[h!]
  \centering
  \begin{tikzpicture}[
    xscale=1, yscale=1,
    every node/.style={align=center},
    font=\small,
    block/.style={rectangle, thick, rounded corners=3pt, minimum height=1cm, text width=4cm},
    layerblock/.style={block, draw=fortissblue!40, fill=fortissblue!5},
    workloadblock/.style={block, draw=fortissblue!70, fill=fortissblue!15},
    largeblock/.style={block, text width=9cm, font=\normalsize, draw=fortissblue!70, fill=fortissblue!25},
    arrow/.style={thick,->,>=Stealth, draw=fortissblue},
    dashedarrow/.style={arrow, dashed}
  ]

  \node[layerblock] (copilot) at (-4,2.5) {General AI Coding Assistants\\\tiny e.g., GitHub Copilot};
  \node[layerblock] (domainai) at (4,2.5) {Domain-specific AI Assistants\\\tiny Tailored to 6G Needs};

  \node[largeblock] (challenge) at (0,1) {\textbf{6G Software Design Challenges}\\\scriptsize Lack of established patterns, limited training data};

  \node[workloadblock] (knowtrans) at (-4,-0.8) {Knowledge Transfer\\\tiny Adaptive Code Generation};
  \node[workloadblock] (onboard) at (4,-0.8) {Developer Support\\\tiny Onboarding, Best Practices};

  \node[largeblock] (impact) at (0,-2.7) {\textbf{Accelerated 6G Software Engineering}\\\scriptsize Faster, Robust, and Secure Development};

  \draw[arrow] (copilot.south) -- (challenge.north west);
  \draw[arrow] (domainai.south) -- (challenge.north east);
  \draw[arrow] (challenge.south west) -- (knowtrans.north);
  \draw[arrow] (challenge.south east) -- (onboard.north);
  \draw[arrow] (knowtrans.south) -- (impact.north west);
  \draw[arrow] (onboard.south) -- (impact.north east);

  \end{tikzpicture}
  \caption{ \textit{The figure illustrates the flow from high-level coding assistants (top) to addressing the core design challenges of 6G (center), enabling knowledge transfer and developer support (middle), which ultimately leads to faster and more robust software development (bottom). Light blue blocks represent general purpose, domain-specific AI coding tools. Darker blocks represent core challenges or system-wide impacts. Medium-light blocks capture development activities supported by AI methods. Solid arrows indicate the flow of influence or functionality from assistants through challenges to implementation outcomes.}
  }
  \label{fig:ai-drivense}
\end{figure}
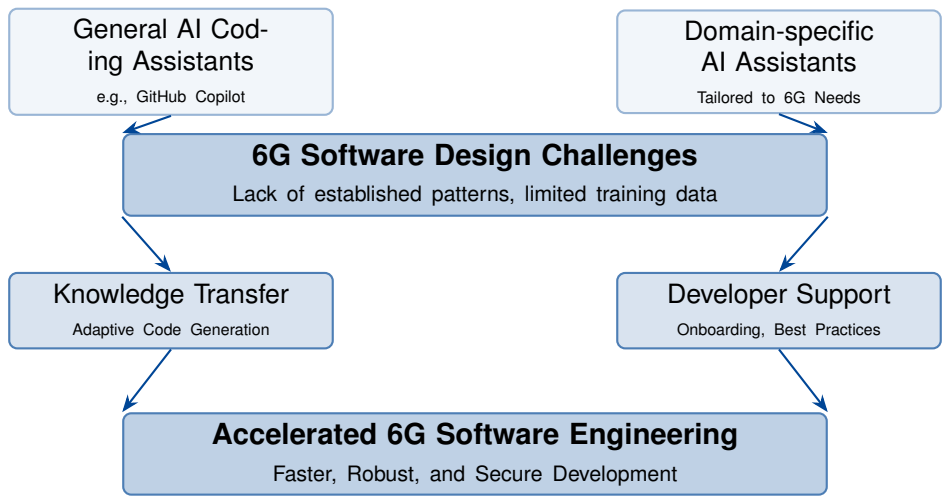

\textbf{fortiss Contributes to This Area By:} Exploring domain-specific AI coding assistants, adaptive code generation, and knowledge transfer methods to improve code quality, reduce development time, and promote best practices in emerging 6G environments, as represented in Figure \ref{fig:ai-drivense}.\hfill


A significant challenge in the development of 6G systems is that developers will initially lack familiarity with effective design patterns and solutions specific to the 6G emerging paradigm. 

Recent advancements in AI-powered coding assistants—such as the GitHub Co-pilot\footnote{https://github.com/features/copilot}
are proven to be useful in accelerating software development by suggesting solutions to common problems, heavily relying on large, existing codebases for training.

With 6G still in an early stages of definition and standardization, limited training data exists in the form of example implementations, protocols, or architectural templates. This scarcity presents 6G as a compelling case study in two key respects. First, it offers a unique opportunity to evaluate the effectiveness of AI coding assistants when applied to domains that are underrepresented in their training data. Second, it highlights the need and potential for developing domain-specific AI coding assistants tailored to 6G. These specialized tools could support developers in navigating the complexity of new 6G architectures, ensuring code quality, encouraging best practices, and reducing time to implementation.

At fortiss, ongoing research into AI-assisted software engineering is exploring how these tools can be adapted to support new and complex domains. This includes research on knowledge transfer in software design, adaptive code generation, and domain-aware AI models.
These models aim to assist developers working in unfamiliar environments. 

Applying these insights to 6G development could not only ease the onboarding process for developers but also promote the creation of robust, efficient, and secure software infrastructures aligned with 6G goals.

\newpage

\section{fortiss in the 6G landscape}
\label{fortiss6G}
\begin{figure}[h!]
  \centering
  \begin{tikzpicture}[
    xscale=1, yscale=1,
    every node/.style={align=center},
    font=\small,
    block/.style={rectangle, thick, rounded corners=3pt, minimum height=1cm, text width=4cm},
    layerblock/.style={block, draw=fortissblue!40, fill=fortissblue!5},
    workloadblock/.style={block, draw=fortissblue!70, fill=fortissblue!15},
    largeblock/.style={block, text width=9cm, font=\normalsize, draw=fortissblue!70, fill=fortissblue!25},
    arrow/.style={thick,->,>=Stealth, draw=fortissblue},
    dashedarrow/.style={arrow, dashed}
  ]
    \node[largeblock] (fortiss) at (0,0) {\textbf{fortiss 6G Contributions}\\Research, Standardization, Collaboration};

    \node[layerblock, above left=0.3cm and 0.3cm of fortiss] (6gia) {\textbf{6G-IA}\\ \tiny \textbf{\textit{AI}} \\ \textbf{\textit{Semantic Communications \\ Edge–Cloud}}};
    \node[layerblock, above right=0.3cm and 0.3cm of fortiss] (networld) {\textbf{\ac{NetworldEurope}}\\Strategic Research and Innovation Agenda, \\ \tiny \textbf{\textit{Softwarization}}\\ };
    \node[layerblock, below left=0.3cm and 0.3cm of fortiss] (one6g) {\textbf{one6G}\\  Performance Indicators \\\tiny \textbf{\textit{Use-Case Analysis}}};
    \node[layerblock, below=3.2cm of fortiss] (conasense) {\textbf{6G CONASENSE}\\ \tiny \textbf{\textit{Localization}}, \\\tiny \textbf{\textit{Sustainability}}, \\\tiny \textbf{\textit{Edge–Cloud}}};
    \node[layerblock, below right=0.3cm and 0.3cm of fortiss] (dia) {\textbf{DIA}\\\tiny \textbf{\textit{Sustainability}}, \\\tiny \textbf{\textit{Architecture}}, \\\tiny \textbf{\textit{Cognition}}};

    \draw[arrow] (fortiss) -- (6gia);
    \draw[arrow] (fortiss) -- (networld);
    \draw[arrow] (fortiss) -- (one6g);
    \draw[arrow] (fortiss) -- (conasense);
    \draw[arrow] (fortiss) -- (dia);
  \end{tikzpicture}
  \caption{\textit{fortiss engagement in international 6G research and standardization initiatives}.}\label{fig:landscape}
\end{figure}
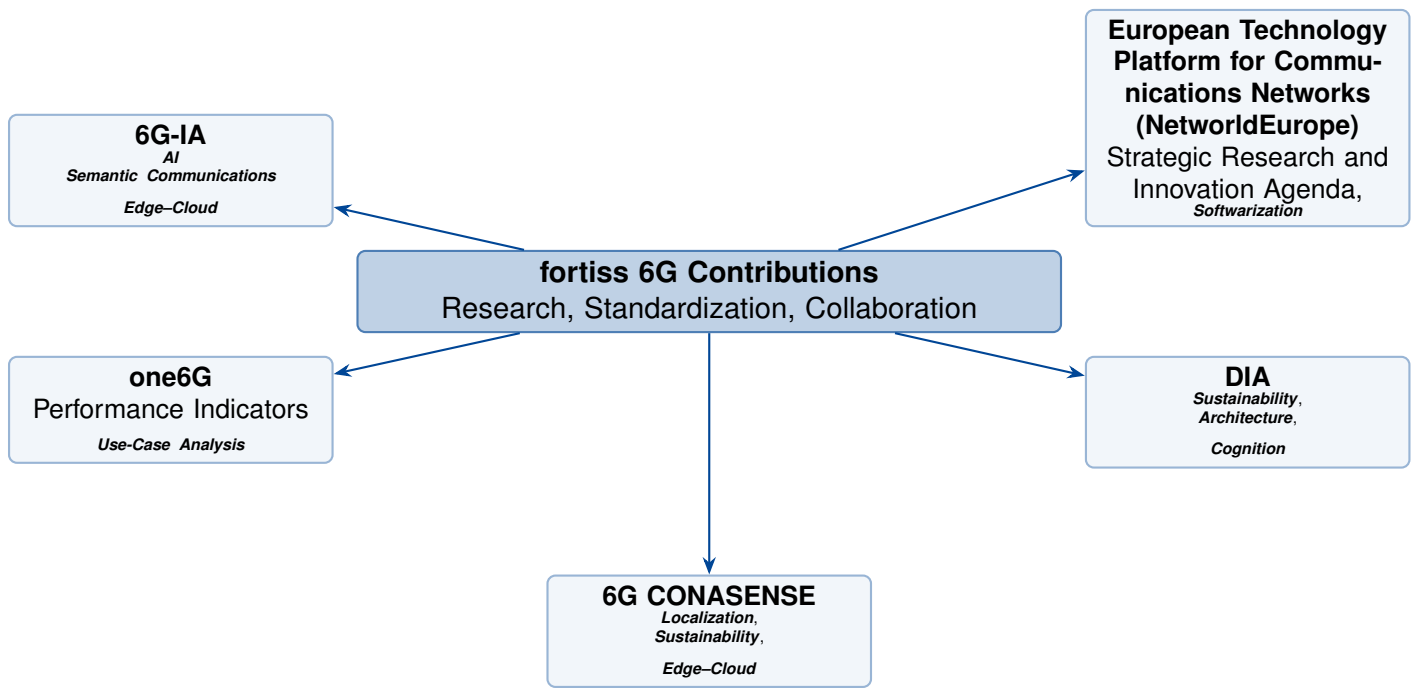

Having outlined the core technologies shaping 6G, such as AI-driven control, Edge–Cloud orchestration, and semantic communications, the next subsections briefly explain fortiss's research initiatives and strategic focus areas that contribute to shaping the future 6G landscape, as summarized in Figure \ref{fig:landscape}.

fortiss is a key contributor to the international 6G landscape (rf. to Figure ), driving innovation through cutting-edge research, active involvement in standardization efforts, and cross-border collaboration. Its work spans critical areas such as sustainability, policy, and next-generation network design—helping to shape 6G systems that are not only intelligent and efficient but also aligned with global societal priorities.
By actively engaging in international research ecosystems, fortiss plays a pivotal role in bridging the gap between technological advancement and real-world needs. Its interdisciplinary approach and commitment to responsible innovation position it as a leader in the transformative journey toward 6G, ensuring that future networks serve both technical excellence and human-centered values.

\subsection{6G-IA: 6G Infrastructure Association}
6G-IA is the leading European industry association dedicated to the development, promotion, and advancement of future communication infrastructures, including 6G networks. It plays a pivotal role in driving research, enabling innovation, and facilitating collaboration across stakeholders within the European Union, while also aligning European efforts with global 6G initiatives.

In \textbf{6G-IA (6G Infrastructure Association)}~\footnote{\url{https://6g-ia.eu/}}, fortiss as Research Member leverages its 6G interdisciplinary expertise in advanced technologies such as AI-driven network management, semantic communications, and Edge–Cloud orchestration, to support the 6G-IA’s mission of fostering research and innovation in next-generation communication systems.

\subsection{Networld-Europe ETP}
NetWorld-Europe is a \textit{\ac{ETP}} dedicated to driving research, innovation, and collaboration in the field of telecommunications and digital infrastructure. It acts as a strategic forum for stakeholders across the \textit{\ac{ICT}} sector, focusing on the advancement of communication networks and their applications, with a particular emphasis on 5G, 6G, and beyond. Originally launched as NetWorld2020, it evolved into NetWorld-Europe to reflect its growing focus on future networks and align with global trends. The platform serves as a key player in shaping the European Union's priorities for research and development in communication technologies, ensuring Europe maintains a competitive edge in the global digital economy.

In \textbf{Networld-Europe}~\footnote{\url{https://www.networldeurope.eu/}}, fortiss has been actively contribute to the mid-term and long-term vision of beyond 5G and 6G, with particular focus on network softwarization and Edge-Cloud orchestration. Regular contributions to the Networld-Europe’s \textit{\ac{SRIA}} in aspects such as decentralized Edge-Cloud orchestration, AI-driven communications.

\subsection{one6G Initiative}

one6G is an international initiative and research consortium dedicated to driving the development, standardization, and deployment of 6G networks and beyond. It acts as a collaborative platform to bring together stakeholders from academia, industry, and government to explore the opportunities and challenges posed by next-generation communication systems. one6G builds on the advancements of 5G while addressing the unique requirements of 6G, including ultra-low latency, energy efficiency, global connectivity, and human-centric designs.

In the one6G initiative~\footnote{\url{https://one6g.org/}} fortiss engages in the analysis and collection of \textit{Key Performance Indicators} (KPIs) for 6G use-cases, and on enabling technologies and system architectures.

\subsection{6G CONASENSE}
The \textbf{6G Communications NAvigation, SEnsing and SErvices (CONASENSE)}\footnote{\url{https://www.conasense.org/}} is a research brainstorming initiative focused on advancing communications technologies, via regular and interdisciplinary brainstorming on 6G main areas of research. It is an open, collaborative and volunteer effort involving researchers, industry experts, and other stakeholders worldwide. Its ultimate goal is to provide the required technical support to develop innovative, human-centric sustainable services backed up by advanced communication systems. 
fortiss leads CONASENSE since 2021, together with the  \textit{CTIF Global Capsule (CGC)} research and transfer institute.
In the core of the CONASENSE vision is a 6G intelligent, heterogeneous wireless system that will provide not only ubiquitous communication but also empower high accuracy localization and high–resolution sensing service. In the envisioned 6G systems, localization, sensing and communication must all coexist, sharing the same time–frequency–spatial resources. However, the support of next generation 6G services must also address interdisciplinary topics, such as energy-awareness by design, integration of user-centric behaviour, joint computation and networking resource management, or adaptation and learning from the far Edge to the Cloud.

\subsection{DIA: Datacom Industry  Association}
\ac{DIA}\footnote{\url{https://datacom-ia.eu/}} aims at fostering discussion and cooperation between 6G data communication and verticals across different competitiveness domains, e.g., Energy, Manufacturing, Transportation, Aerospace. The key focus is on strategic research and innovation topics such as security and privacy; routing and addressing; control plane; telemetry. DIA aims at forming  a one-stop community for \textit{Information and Communication Technology (ICT)} experts experts to discuss and to cooperate with vertical experts, to agree on what new researches, standards and policies are needed, and collaborate to realize them.

fortiss is a DIA Founding Member and is also in the DIA Board, participating actively in the definition of DIA, as well as in research concerning an advanced cognitive plane, routing, and orchestration.

\newpage
\section{Vision towards 2030}
\label{6Gfutureinfortiss}

6G is currently being envisioned as a connected system of systems, which is expected to bring a paradigm shift in terms of the societal and technological environments. 

In contrast to the former "Gs", 6G represents a fundamental transformation in wireless technology, seamlessly integrating communications, sensing, computation and AI to deliver human-centric, immersive experiences. At the core of this vision is a future decentralized (de-aggregated) AI-enabled architecture, where sustainability principles are expected to be integrated by design, and where NTN and terrestrial communications are the backbone for ubiquitous, smart, open connectivity.

The fortiss vision considers that AI-driven decision-making is expected to enable dynamic, human-centric orchestration of computational, network resources and and context-awareness across the entire continuum, from the far edge to the cloud. This intelligent orchestration will support real-time infrastructure optimization while also addressing energy efficiency and contributing to a reduced overall carbon footprint.

Emerging XR-based and multi-sensory applications, combined with the large-scale deployment of IoT devices across both personal and industrial domains, will drive significant advancements in digitization and automation. These developments will underpin future cognitive and autonomous systems, digital twins, and advanced healthcare solutions. With the number of IoT devices projected to surpass 32 billion by 2030, existing communication models will face increasing strain.

To address this, novel paradigms such as semantic communications—and eventually quantum communications—will be essential. These approaches aim to reduce data redundancy, optimize bandwidth usage, and enable context-aware, mission-critical transmissions that align with the real-time demands of complex, interconnected systems.

At its core, the evolution of 6G will be guided by human-centric principles, with a strong focus on \textbf{inclusivity}, \textbf{accessibility}, and \textbf{enhanced QoE} across diverse user groups and usage contexts. Realizing this vision demands more than just advanced technologies: \textbf{it requires a deliberate and sustained commitment to designing systems that align with human values, needs, and capabilities, ensuring that 6G serves as an enabler of equitable and meaningful digital experiences}.

In this context, cross-disciplinary collaboration is essential, involving experts from telecommunications, AI, social sciences, and sustainability domains to ensure ethical, secure, and inclusive 6G development. As global initiatives continue to shape the 6G landscape, fortiss will remain at the forefront of this evolution, contributing to intelligent, secure, and human-centric 6G services that redefine digital interactions.
fortiss envisions 6G as a \textbf{cognitive, and decentralized networking system of systems} that seamlessly integrates communication, computation, and data orchestration across an advanced Edge-Cloud continuum. By leveraging AI-driven network intelligence and frameworks like \ac{CODECO}\footnote{https://gitlab.eclipse.org/eclipse-research-labs/codeco-project/}, 6G will enable dynamic, self-optimizing, and scalable resource management. This approach will support AI-native networks, capable of making autonomous decisions, optimizing traffic flow, and ensuring \textit{ultra-reliable low-latency communication (URLLC)} for emerging applications in smart cities, autonomous systems, energy, health, and industrial automation.

\textbf{Greenness by design} is central to fortiss' 6G research. Future networks must minimize energy consumption while maintaining high-performance connectivity. fortiss is actively exploring cross-layer energy-efficient network designs, where decentralized computing, federated AI, decentralized learning, and dynamic energy-aware orchestration will be key enablers of green 6G networks. Additionally, fortiss contributes to sustainable software development, lifecycle management, and circular economy initiatives to reduce the environmental footprint of communication technologies.

Finally, fortiss is committed to \textbf{shaping the global 6G} ecosystem through \textbf{active participation in standardization efforts and cross-disciplinary collaboration}. Engaging with 6G-IA, NetworldEurope, one6G, and CONASENSE, fortiss aligns its research with international 6G roadmaps, ensuring that 6G advancements support ethical AI, secure digital infrastructures, and future-proof communication systems. By integrating AI-driven networking, decentralized intelligence, and sustainable engineering, Fortiss contributes to a resilient, intelligent, and adaptive 6G era that will redefine how humans and cyber-physical systems interact in 2030 and beyond.
In this context, the contributions towards the 6G 2030 vision being researched by fortiss are summarized in Table \ref{tab:research_pillars}.

\begin{table}[!ht]
    \centering
      \caption{\textit{Key Research Pillars in fortiss' 6G Research.}}
    \label{tab:research_pillars}
    \scriptsize
    \renewcommand{\arraystretch}{1.3}
   \resizebox{\textwidth}{!}{%

    \begin{tabular}{|p{3cm}|p{3cm}|p{5cm}|p{4cm}|}
      \hline
\textbf{Pillar} & \textbf{Challenge} & \textbf{Research Contribution} & \textbf{Expected Impact} \\
        \hline
        \textbf{Edge-Cloud Orchestration} & Traditional cloud-centric architectures introduce latency, inefficiency, and privacy concerns for 6G applications. & - Development of AI-driven orchestration frameworks (e.g., CODECO) for dynamic and adaptive resource allocation across Edge, Cloud, and network layers. - Exploration of decentralized computing models to reduce central bottlenecks. & Enhanced performance, reliability, and real-time service provisioning in 6G networks. \\

        \hline
        \textbf{AI-as-a-Service} & 
        AI in current networks remains largely centralized and lacks the adaptability required for the 6G Edge-Cloud continuum. & 
        - Development of AI-as-a-Service models for scalable, on-demand AI-driven network management. 
        - Exploration of decentralized learning approaches (e.g., Swarm Learning, Federated Learning, MARL) for effective 6G-AI integration. & 
        AI-native 6G networks that are self-learning, self-optimizing, and highly resilient. \\
        \hline
        \textbf{Green Network Architectures} & 
        Increasing network energy consumption conflicts with sustainability and carbon reduction targets. & 
        - AI-driven energy-aware network resource allocation. 
        - Development of cross-layer energy-efficient models integrating IoT, Edge-Cloud, and AI. & 
        Green 6G networks reducing carbon footprint while maintaining high-performance connectivity. \\
        \hline
        \textbf{Semantic Communications \& Context-Aware Networking} & 
        Traditional bit-based communication is inefficient for 6G applications requiring intelligent data transmission. & 
        - Development of semantic-aware communication architectures that prioritize relevant data over raw transmission. 
        - Implementation of context-aware IoT and intelligent data filtering to reduce network congestion. & 
        Improved bandwidth efficiency, lower latency, and better resource utilization. \\
        \hline
        \textbf{Network as a Sensor} & 
        Conventional networks focus only on data transmission, lacking sensing capabilities. & 
        - Development of 6G localization, radar-based communication, and JCAS systems for environmental awareness. 
        - Research on JCAS for applications in smart cities, autonomous systems, and human-machine interaction. & 
        AI-driven real-time environmental awareness enabling advanced 6G services. \\
        \hline
        \textbf{Human-centered 6G design} & 
        Demands on aligning advanced AI-driven technologies with human needs, ensuring trust, transparency, inclusivity, privacy and usability. & 
        - Developing explainable models ensuring AI-driven decisions transparent and trustworthy . 
        - Designing systems that can adapt to user preferences, needs, and cognitive models. 
        - Human-in-the-Loop integration enables scalable human involvement in AI decision-making of 6G.  & 
        Enable human-centric design of 6G. \\
        \hline
        \textbf{6G Standardization and Policies} & 
        6G standards are still evolving, requiring coordinated research, industry, and policy efforts. & 
        - fortiss contributes to global 6G standardization via 6G-IA, DIA, NetworldEurope, one6G, CONASENSE, and other initiatives. 
        - Focus on cross-sector collaboration to align 6G with industry and sustainability goals. & 
        Bring awareness and assist in a human-centric, energy-efficient, and resilient design of 6G. \\
        \hline
    \end{tabular}
  }  
\end{table}

\newpage

\section{Summary}
\label{Conclusions}
The evolution of 6G marks a paradigm shift beyond previous generations, placing a strong emphasis on human-centric design, AI-driven intelligence, and the seamless integration of cyber-physical systems. As IoT connectivity expands and immersive digital experiences become more widespread, 6G will demand a robust, scalable, and interdisciplinary approach to innovation.

fortiss is actively contributing to this transformation by advancing Edge–Cloud continuum architectures, developing AI-powered network optimization strategies, and promoting user-centric design methodologies. In addition to its technical contributions, fortiss plays a vital role in standardization efforts and cross-sector collaboration, ensuring that 6G development is aligned with the needs of industry, academia, and society at large.

Key takeaways from fortiss's 6G research efforts highlight the need for decentralized, AI-driven network management, sustainable and intelligent infrastructure, and real-world use cases spanning smart cities, industrial automation, and energy systems. The institute’s commitment to interdisciplinary research strengthens its position as a key player in shaping the 6G ecosystem. As global initiatives push toward 6G standardization and pilot deployments by 2030, fortiss remains at the forefront, driving innovation that ensures 6G serves as a transformative force for connectivity, intelligence, and societal advancement.
\textbf{The current ongoing research emphasizes the following aspects:}
\begin{itemize}
    \item \textbf{6G is more than faster speeds and high throughput} — our vision emphasize intelligence, automation, and sustainability.
    \item \textbf{AI-driven networking and decentralization} will enable self-optimizing, scalable networks.
    \item \textbf{Energy efficiency is a priority}, ensuring sustainable and carbon-neutral infrastructure.
    \item \textbf{6G will empower human-centric services} through large-scale integration of intelligent sensing, ubiquitous IoT, and adaptive interfaces to deliver personalized, context-aware experience.
    
\end{itemize}

\newpage
\section{Call to Action: Shaping a Human-Centered 6G Future}

The development of 6G is not just a technological endeavor—it is a societal one. To realize a truly inclusive, sustainable, and intelligent communication infrastructure, collaboration across disciplines and sectors is essential.

We invite:
\begin{itemize}
    \item \textbf{Academic researchers} to engage with fortiss in exploring interdisciplinary challenges, from AI ethics to edge intelligence.
    \item \textbf{Industry partners} to co-develop testbeds and applications that stress-test 6G concepts in real-world scenarios.
    \item \textbf{Policymakers and standards bodies} to align regulatory frameworks with the human-centric, privacy-preserving principles envisioned for 6G.
    \item \textbf{Developers and innovators} to contribute to open tools and coding platforms that accelerate 6G adoption.

\end{itemize}

Interested partners can engage via: (1) Joint Horizon Europe project proposals, (2) fortiss-hosted 6G initiatives, (3) contribution to experimental deployments based on open-source software.

\newpage

\section*{Contributors}
We thank all of the fortiss colleagues that have contributed to the 6G white paper version 1.0.

\begin{table}[htpb]
\begin{tabularx}{\textwidth}{|X|X|}
\hline
\textbf{Contributor Name} & \textbf{Position} \\
\hline
Rute C. Sofia & IIoT Head (Ed.) \\
Hao Shen & Machine Learning Head\\
Yuanting Liu & Human-computer Engineering Head \\
Severin Kacianka & Center for Code Excellence Head \\
Holger Pfeifer & CEO \\
\hline
\end{tabularx}
\label{tab:contributors}
\end{table}



%
\bibliographystyle{IEEEtran}
\bibliography{6Greferences}

\begin{thebibliography}{10}
\providecommand{\url}[1]{#1}
\csname url@samestyle\endcsname
\providecommand{\newblock}{\relax}
\providecommand{\bibinfo}[2]{#2}
\providecommand{\BIBentrySTDinterwordspacing}{\spaceskip=0pt\relax}
\providecommand{\BIBentryALTinterwordstretchfactor}{4}
\providecommand{\BIBentryALTinterwordspacing}{\spaceskip=\fontdimen2\font plus
\BIBentryALTinterwordstretchfactor\fontdimen3\font minus
  \fontdimen4\font\relax}
\providecommand{\BIBforeignlanguage}[2]{{%
\expandafter\ifx\csname l@#1\endcsname\relax
\typeout{** WARNING: IEEEtran.bst: No hyphenation pattern has been}%
\typeout{** loaded for the language `#1'. Using the pattern for}%
\typeout{** the default language instead.}%
\else
\language=\csname l@#1\endcsname
\fi
#2}}
\providecommand{\BIBdecl}{\relax}
\BIBdecl

\bibitem{5g_evolution}
U.~B. Shukurillaevich, R.~O. Sattorivich, and R.~U. Amrillojonovich, ``{5G
  technology evolution},'' in \emph{2019 International Conference on
  Information Science and Communications Technologies (ICISCT)}.\hskip 1em plus
  0.5em minus 0.4em\relax IEEE, 2019, pp. 1--5.

\bibitem{industryautomation}
M.~Attaran, ``{The impact of 5G on the evolution of intelligent automation and
  industry digitization},'' \emph{Journal of ambient intelligence and humanized
  computing}, vol.~14, no.~5, pp. 5977--5993, 2023.

\bibitem{pennanen20246g}
H.~Pennanen, T.~H{\"a}nninen, O.~Tervo, A.~T{\"o}lli, and M.~Latva-aho, ``{6G:
  The Intelligent Network of Everything--A Comprehensive Vision, Survey, and
  Tutorial},'' \emph{arXiv e-prints}, pp. arXiv--2407, 2024.

\bibitem{alsamhi2024multisensory}
M.~H. Alsamhi, A.~Hawbani, S.~Kumar, and S.~H. Alsamhi, ``{Multisensory
  metaverse-6G: A new paradigm of commerce and education},'' \emph{IEEE
  Access}, 2024.

\bibitem{liu2025ituvisionframework6g}
\BIBentryALTinterwordspacing
R.~Liu, L.~Zhang, R.~Y.-N. Li, and M.~D. Renzo, ``{The ITU Vision and Framework
  for 6G: Scenarios, Capabilities and Enablers},'' 2025. [Online]. Available:
  \url{https://arxiv.org/abs/2305.13887}
\BIBentrySTDinterwordspacing

\bibitem{butt2024ambient}
M.~M. Butt, N.~R. Mangalvedhe, N.~K. Pratas, J.~Harrebek, J.~Kimionis,
  M.~Tayyab, O.-E. Barbu, R.~Ratasuk, and B.~Vejlgaard, ``{Ambient IoT: A
  missing link in 3GPP IoT devices landscape},'' \emph{IEEE Internet of Things
  magazine}, vol.~7, no.~2, pp. 85--92, 2024.

\bibitem{veedu2022toward}
S.~N.~K. Veedu, M.~Mozaffari, A.~H{\"o}glund, E.~A. Yavuz, T.~Tirronen,
  J.~Bergman, and Y.-P.~E. Wang, ``{Toward smaller and lower-cost 5G devices
  with longer battery life: An overview of 3GPP release 17 RedCap},''
  \emph{IEEE Communications Standards Magazine}, vol.~6, no.~3, pp. 84--90,
  2022.

\bibitem{banafaa20236g}
M.~Banafaa, I.~Shayea, J.~Din, M.~H. Azmi, A.~Alashbi, Y.~I. Daradkeh, and
  A.~Alhammadi, ``{6G mobile communication technology: Requirements, targets,
  applications, challenges, advantages, and opportunities},'' \emph{Alexandria
  Engineering Journal}, vol.~64, pp. 245--274, 2023.

\bibitem{hu2024survey}
H.~Hu, Z.~Wang, X.~Zhao, R.~Li, A.~Li, Y.~Si, J.~Wang, T.~Zhou, and T.~Xu, ``{A
  survey on brain-computer interface-inspired communications: opportunities and
  challenges},'' \emph{IEEE Communications Surveys \& Tutorials}, 2024.

\bibitem{sofia2024framework}
R.~C. Sofia, J.~Salomon, S.~Ferlin-Reiter, L.~Garc{\'e}s-Erice, P.~Urbanetz,
  H.~Mueller, R.~Touma, A.~Espinosa, L.~M. Contreras, V.~Theodorou
  \emph{et~al.}, ``{A framework for cognitive, decentralized container
  orchestration},'' \emph{IEEE Access}, 2024.

\bibitem{c2024shaping}
R.~C~Sofia and J.~Soldatos, \emph{{Shaping the Future of IoT with Edge
  Intelligence: How Edge Computing Enables the Next Generation of IoT
  Applications}}.\hskip 1em plus 0.5em minus 0.4em\relax Taylor \& Francis,
  2024.

\bibitem{sofia2025green}
\BIBentryALTinterwordspacing
R.~Sofia and D.~Ali, ``{Energy-aware Differentiated Services (EA-DS)},''
  Internet Engineering Task Force, Internet-Draft, IETF GREEN proposal
  draft-sofia-green-energy-aware-diffserv-00, July 2025, active. [Online].
  Available:
  \url{https://datatracker.ietf.org/doc/draft-sofia-green-energy-aware-diffserv/}
\BIBentrySTDinterwordspacing

\bibitem{lins2021artificial}
S.~Lins, K.~D. Pandl, H.~Teigeler, S.~Thiebes, C.~Bayer, and A.~Sunyaev,
  ``{Artificial intelligence as a service: classification and research
  directions},'' \emph{Business \& Information Systems Engineering}, vol.~63,
  pp. 441--456, 2021.

\bibitem{wang2024distributed}
Y.~Wang, Z.~Tian, X.~Fan, Z.~Cai, C.~Nowzari, and K.~Zeng, ``{Distributed swarm
  learning for edge internet of things},'' \emph{IEEE Communications Magazine},
  2024.

\bibitem{gao2020end}
Y.~Gao, M.~Kim, S.~Abuadbba, Y.~Kim, C.~Thapa, K.~Kim, S.~A. Camtepe, H.~Kim,
  and S.~Nepal, ``{End-to-end evaluation of federated learning and split
  learning for internet of things},'' \emph{arXiv preprint arXiv:2003.13376},
  2020.

\bibitem{xiao2024llm}
Z.~Xiao, C.~Ye, Y.~Hu, H.~Yuan, Y.~Huang, Y.~Feng, L.~Cai, and J.~Chang, ``{LLM
  Agents as 6G Orchestrator: A Paradigm for Task-Oriented Physical-Layer
  Automation},'' \emph{arXiv preprint arXiv:2410.03688}, 2024.

\bibitem{lin2023pushing}
Z.~Lin, G.~Qu, Q.~Chen, X.~Chen, Z.~Chen, and K.~Huang, ``{Pushing large
  language models to the 6g edge: Vision, challenges, and opportunities},''
  \emph{arXiv preprint arXiv:2309.16739}, 2023.

\end{thebibliography}

%


\newpage
\section*{List of Acronyms}

\begin{acronym}[ACRONYM] 
\acro{6G} {Sixth Generation Mobile Communication}
\acro{5G} {Fifth Generation Mobile Communication}
\acro{AI} {Artificial Intelligence}
\acro{A-IoT} {Ambient Internet of Things}
\acro{AR} {Augmented Reality}
\acro{BCI} {Brain-Computer Interfaces}
\acro{CODECO} {Cognitive, Decentralized Edge-Cloud Orchestration}
\acro{DIA} {Datacom Industry Association}
\acro{ETP} {European Technology Platform}
\acro{HAP} {High-Altitude Platform}
\acro{ICT} {Information and Communication Technology}
\acro{IIoT} {Industrial Internet of Things}
\acro{IMT} {International Mobile Telecommunications}
\acro{IoT} {Internet of Things}
\acro{ISAC} {Integrated Sensing and Communications}
\acro{ITU} {International Telecommunication Union}
\acro{ITU-R} {International Telecommunication Union Radiocommunication Sector}
\acro{JCAS} {Joint Communication and Sensing}
\acro{MARL} {Multi-Agent Reinforcement Learning}
\acro{ML} {Machine Learning}
\acro{NetworldEurope} {European Technology Platform for Communications Networks}
\acro{NTN} {Non-Terrestrial Networks}
\acro{PQC} {Post-Quantum Cryptography}
\acro{QoE} {Quality of Experience}
\acro{QoS} {Quality of Service}
\acro{RFID} {Radio-Frequency Identification}
\acro{RL} {Reinforcement Learning}
\acro{SEAL} {Service Enabler Architecture Layer for Verticals}
\acro{SLAM} {Simultaneous Localization and Mapping}
\acro{SRIA} {Strategic Research and Innovation Agenda}
\acro{SSB} {Synchronization Signal Block}
\acro{UE} {User Equipment}
\acro{URLLC} {Ultra-Reliable Low-Latency Communication}
\acro{UAV} {Unmanned Aerial Vehicle}
\acro{VR} {Virtual Reality}
\acro{XR} {Extended Reality}
\end{acronym}

\end{document}